\documentclass[a4paper,12pt]{article}

\usepackage{amsmath,amssymb,amsfonts}
\allowdisplaybreaks
\usepackage{graphicx}
\usepackage{color}
\makeatletter
\@addtoreset{equation}{section}
\renewcommand{\theequation}{\thesection.\@arabic\c@equation}
\makeatother
\usepackage{hyperref}
\usepackage{cite}
\usepackage{caption}
\definecolor{red}{rgb}{1,0,0}
\definecolor{green}{rgb}{0,1,0}
\definecolor{blue}{rgb}{0,0,1}
\definecolor{darkblue}{rgb}{0,0,0.5}
\definecolor{lightblue}{rgb}{.5,.5,1}
\definecolor{lightgray}{gray}{.87}
\definecolor{Dark}{gray}{.20}
\definecolor{pink}{rgb}{.95,0.82,0.92}
\definecolor{yellow}{rgb}{1,1,0}
\definecolor{lightyellow}{rgb}{1,1,.5}
\definecolor{purple}{rgb}{0.7,0,0.85}
\definecolor{darkgreen}{rgb}{0,0.5,0}
\definecolor{darkpurple}{rgb}{0.4,0,0.4}
\definecolor{orange}{rgb}{0.8,0.2,0.2}
\def \be {\begin{equation}}
\def \ee {\end{equation}}
\def \bea {\begin{align}}
\def \eea {\end{align}}
\def \nn {\nonumber}

\def \rr {\raise.35ex\hbox{\small $\prime$}\kern-.17em{\mbox{\large $\imath$}}}
\def \del {\partial}
\def \dels {\partial\kern-.5em / \kern.5em}
\def \As {{A\kern-.5em / \kern.5em}}
\def \Ds {D\kern-.7em / \kern.5em}

\def \dag {\dagger}

\def \d {\delta}

\def \lam {\lambda}

\def \om {\omega}

\def \th {\theta}

\newcommand{\hide}[1]{}

\newcommand{\explanation}[1]{}

\pdfoutput=1

\begin{document}

\pagestyle{plain}


\begin{titlepage}
\vspace*{-10mm}   
\baselineskip 10pt   
\begin{flushright}   
\begin{tabular}{r} 
\end{tabular}   
\end{flushright}   
\baselineskip 24pt   
\vglue 10mm

\begin{center}

\noindent
\textbf{\LARGE
Firewall From Effective Field Theory
}
\vskip20mm
\baselineskip 20pt

\renewcommand{\thefootnote}{\fnsymbol{footnote}}

{\large
Pei-Ming~Ho${}^a$
\footnote[1]{pmho@phys.ntu.edu.tw},
Yuki~Yokokura$^b$
\footnote[2]{yuki.yokokura@riken.jp}
}

\renewcommand{\thefootnote}{\arabic{footnote}}

\vskip5mm

{\it  
${}^{a}$
Department of Physics and Center for Theoretical Physics, \\
National Taiwan University, Taipei 106, Taiwan,
R.O.C. 
\\
${}^{b}$
iTHEMS Program, RIKEN, Wako, Saitama 351-0198, Japan
}

\vskip 25mm
\begin{abstract}

For an effective field theory in the background of 
an evaporating black hole with spherical symmetry,
we consider non-renormalizable interactions
and their relevance to physical effects. 
The background geometry is determined 
by the semi-classical Einstein equation
for an uneventful horizon 
where the vacuum energy-momentum tensor is small
for freely falling observers.
Surprisingly,
after Hawking radiation appears,
the transition amplitude from the Unruh vacuum to certain multi-particle states
grows exponentially with time for a class of higher-derivative operators
after the collapsing matter enters the near-horizon region,
despite the absence of large curvature invariants.
Within the scrambling time,
the uneventful horizon transitions towards a firewall,
and eventually the effective field theory breaks down.

\end{abstract}
\end{center}

\end{titlepage}

\pagestyle{plain}

\baselineskip 18pt

\setcounter{page}{1}
\setcounter{footnote}{0}
\setcounter{section}{0}


\newpage

\section{Introduction}\label{introduction}

The information loss paradox \cite{Hawking:1976ra,Mathur:2009hf,Marolf:2017jkr}
has been puzzling theoretical physicists
since the discovery of Hawking radiation \cite{Hawking-Radiation}.
Nowadays, most people,
including Hawking \cite{Hawking-Weather},
believe that there is no information loss
at least for a consistent theory of quantum gravity such as string theory.
But a persisting outstanding question is
how string theory (or any theory of quantum gravity) ever becomes relevant
during the evaporation of black holes.
\footnote{
Other outstanding questions about the paradox 
include whether Hawking radiation is thermal,
and how its entanglement entropy should be computed.
There is significant recent progress in these directions
\cite{Saini:2015dea,Zhang-Hawking,AdS-Entropy}.
}
That is,
how does the low-energy effective theory break down
in the absence of high-energy events
\footnote{A ``high-energy event'' refers to a physical observable
at an energy scale higher than the cutoff energy
of the low-energy effective theory.}
\cite{Mathur:2009hf}?

If there is no high-energy event around the horizon,
the effective theory
is expected to be a good approximation.
But it is incapable of describing the transfer of 
the complete information 
inside arbitrary collapsing matter into the outgoing radiation. 
For example, 
the information hidden inside a nucleus in free fall
cannot be retrieved unless there are events (e.g. scatterings)
above the scale of the QCD binding energy \footnote{
There is no clear inconsistency in a unitary evaporation 
without high-energy events \cite{Hutchinson:2013kka}
if there are no small particles like nuclei.
But we will show below that a firewall
still arises under general assumptions.
}.
This conflict between an uneventful horizon and unitarity
has been emphasized in Refs.\cite{Mathur:2009hf,firewall}
and it has motivated the proposals of fuzzballs \cite{FuzzBall}
and firewalls \cite{firewall,firewall-B}.

It has been shown \cite{Dodelson:2015toa} that
the effective field theory of string theory breaks down
in the near-horizon regime due to stringy effects.
The mechanism involved is not directly related to the one studied here.
More importantly,
we emphasize that,
to resolve the information loss paradox,
we must identify an abnormal process in the low-energy effective theory
as a warning or signal that the low-energy effective  theory is breaking down.
Otherwise,
how can we be sure that
the application of low-energy effective theories to
any problem at arbitrarily low energies
would not also break down unexpectedly?

In the modern interpretation of quantum field theories
(see e.g. \S 12.3 of Ref.\cite{Weinberg:1995mt}),
the effective Lagrangian (see eq.\eqref{action-4D} below)
includes all higher-dimensional local operators
which are normally assumed to be negligible at low energies
because they are suppressed by powers of $1/M_p$,
where $M_p$ is the Planck mass (or the cut-off energy).
It is well known that,
when there are Planck-scale curvatures,
the higher-dimensional terms cannot be ignored,
and the effective field-theoretic description fails.
However, no rigorous proof has been given to show
that a non-trivial spacetime geometry without large curvature
cannot introduce significant physical effects through these non-renormalizable interactions.
In this paper, 
we show that there are indeed higher-dimensional interactions with large physical effects 
in the near-horizon region where the curvature is small, 
and that this eventually leads to the formation of a firewall
and the breakdown of the effective field theory
within the time scale of the so-called ``scrambling time'' \cite{Sekino:2008he}.

In the derivation of the firewall,
we assume that the effective-field-theoretic derivation of Hawking radiation is valid.
(This assumes the presence of certain high-frequency modes in the quantum fluctuation.)
Hence,
strictly speaking,
the conclusion is that our understanding of the Hawking radiation is 
incompatible with the uneventful horizon
over a time scale longer than the scrambling time.

We construct in Sec.\ref{geometry} the spacetime geometry 
for a dynamical black hole with an uneventful horizon, 
including the back-reaction of the vacuum energy-momentum tensor. 
``Uneventful" means that 
there is no high-energy event 
and the energy-momentum tensor is small for freely falling observers
comoving with the collapsing matter.
We show in Sec.\ref{effectivefieldtheory} that,
after the collapsing matter enters the near-horizon region,
certain (higher-dimensional) higher-derivative interaction terms, 
which are naively suppressed by powers of $1/M_p^{2n}$ (for $n > 1$),
lead to an exponentially growing probability of
transition to certain multi-particle states
from the Unruh vacuum
within the time scale
$\Delta t \sim \mathcal{O}\left(\frac{1}{n}a\log \frac{a}{\ell_p}\right)$
for large $n$.
Here, 
$t$ is the time for distant observers, 
$a$ is the Schwarzschild radius of the black hole, 
and $\ell_p = 1/M_p$ is the Planck length.
The created particles
have high energies as a firewall for freely falling observers.
Eventually, the effective field theory breaks down.
We conclude in Sec.\ref{conclusion}
with comments on potential implications of our results.

We use the convention $\hbar = c = 1$ in this paper.

\section{Back-reacted geometry}
\label{geometry}
A hint at the invalidity of low-energy effective theories around the horizon
was the recent finding \cite{HMY-1,ShortDistance} that,
until the black hole is evaporated to a tiny fraction of its initial mass,
the proper distance between the trapping horizon
and the surface of the collapsing matter is at most a few Planck lengths,
although the curvature is still small. 
Such a near-horizon geometry of the dynamical black hole
has a Planck-scale nature which is not characterized 
by the curvature invariants. 
In this section, we describe the geometry around the near-horizon region 
by reviewing and extending the results of Refs.\cite{HMY-1,ShortDistance}.

We consider the gravitational collapse of a null matter
of finite thickness
from the infinite past.
The spacetime geometry is determined by 
the expectation value $\langle T_{\mu\nu} \rangle$ of the energy-momentum tensor
through the semi-classical Einstein equation
\begin{align}
G_{\mu\nu} = \kappa \langle T_{\mu\nu} \rangle,
\label{SCEE}
\end{align}
where $\kappa \equiv 8\pi G_N$.

Assuming spherical symmetry,
the metric can be written in the form
\be
ds^2 = - C(u, v) du dv + r^2(u, v) d\Omega^2.
\label{metric}
\ee
We shall consider an asymptotically flat spacetime
and adopt the convention that $C(u, v) \rightarrow 1$
at large distances. 

In the classical limit, 
$\langle T_{\mu\nu} \rangle = 0$
for the space outside the matter,
and the geometry is described by the Schwarzschild metric:
\begin{align}
C(u, v) &= 1 - \frac{a}{r},
\label{C-Schwarzschild}
\\
\frac{\del r}{\del u} &= - \frac{\del r}{\del v} = - \frac{1}{2}\left(1 - \frac{a}{r}\right),
\label{Sch-drdu}
\end{align}
where $a$ is the Schwarzschild radius.

The vacuum energy-momentum tensor $\langle T_{\mu\nu} \rangle$
leads to a quantum correction to this solution via eq.\eqref{SCEE}.  
While the classical solution has a curvature tensor $\sim \mathcal{O}(1/a^2)$,  
the vacuum energy-momentum tensor is
$\kappa \langle T_{\mu\nu} \rangle \sim \mathcal{O}(\ell_p^2/a^4)$
(see eqs.\eqref{Tuu} -- \eqref{Tthth} below). 
Therefore, in the Einstein equation \eqref{SCEE},
we can take $\ell_p^2/a^2$ as the dimensionless parameter
to treat the quantum correction perturbatively
well outside the horizon where $C(u, v) \gg \mathcal{O}(\ell_p^2/a^2)$.
Such treatment has been widely applied to
the study of black-hole geometry in the literature.
On the other hand,
the geometry close to the horizon could be modified more significantly.

Following recent progresses \cite{Ho:2019kte,HMY-1,ShortDistance},
we give in this section the approximate solution
to the semi-classical Einstein equation in the near-horizon region
for an adiabatic process.
It is characterized by two (generalized)
time-dependent Schwarzschild radii $a(u)$ and $\bar{a}(v)$
(see eq.\eqref{def_aa} for their definitions).\footnote{
The solution is consistent with previous studies on special cases
\cite{Ho:2017joh,Ho:2017vgi,Ho:2018jkm,Ho:2018fwq,HMY-1}.
}
Both $a(u)$ and $\bar{a}(v)$ agree with the classical Schwarzschild radius $a$
in the limit $\ell_p/a \rightarrow 0$.

\subsection{Near-horizon region and uneventful condition} 
\begin{figure}
\center
\includegraphics[scale=0.5,bb=0 0 500 250]{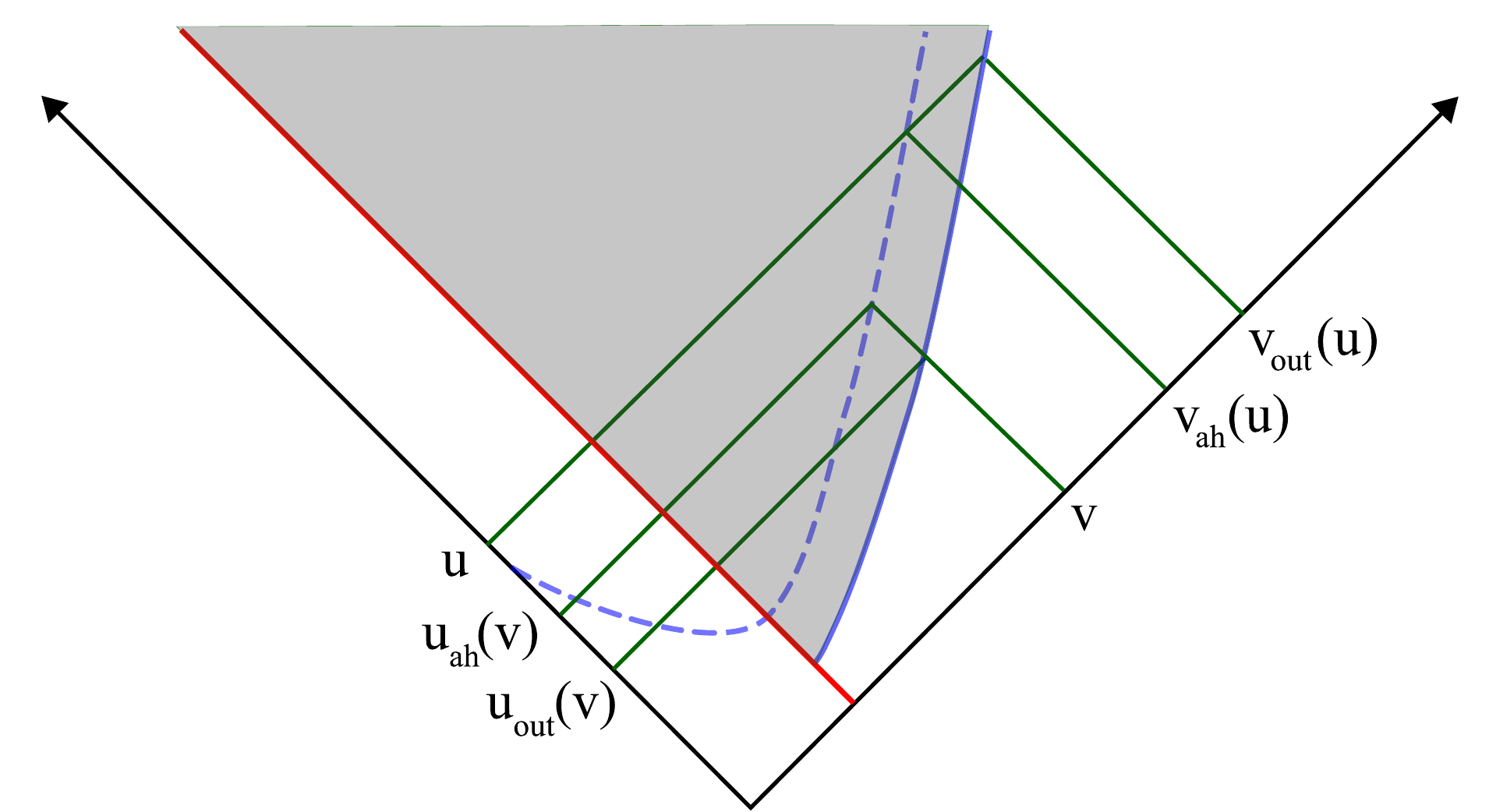}
\caption{\small
The solid blue curve is the outer boundary of the near-horizon region.
The red straight line represents the surface of the null collapsing matter,
and the shaded area the near-horizon region.
The dash blue curve is the trapping horizon.
}
\label{UV}
\vskip1em
\end{figure}

We start by reviewing the definition of the {\em near-horizon region}.
Roughly speaking,
it is defined to be the region near and inside the trapping horizon,
but outside the collapsing matter \cite{ShortDistance}.
The surface of the collapsing matter is the inner boundary of the near-horizon region.
The outer boundary is slightly outside the trapping horizon
where the Schwarzschild approximation is valid. 
We will restrict our consideration to the early stage of black-hole evaporation
when the trapping horizon is timelike in the near-horizon region.
(See Fig.\ref{UV}.)

The definition of the outer boundary of the near-horizon region is clearly not unique.
Nevertheless, since the quantum correction is small when
$C(u, v) \gg \mathcal{O}(\ell_p^2/a^2)$,
or equivalently,
when $r(u, v) - a \gg \ell_p^2/a$
according to eq.\eqref{C-Schwarzschild},
it is reasonable to define it by the condition
\begin{align}
r(u_{out}(v), v) - \bar{a}(v) = \frac{N\ell_p^2}{\bar{a}(v)}
\qquad (N \gg 1),
\label{outer-boundary-condition}
\end{align}
where $u_{out}(v)$ is the $u$-coordinate
of the outer-boundary of the near-horizon region
for a given value of $v$.
\footnote{
It is equally natural to use the condition
$r(u, v_{out}(u)) - a(u) = N\ell_p^2/a(u)$
instead of eq.\eqref{outer-boundary-condition}.
This different choice would not make any essential difference 
in the discussion below.
}
The number $N$ should be so large that 
the Schwarzschild metric with the Schwarzschild radius $\bar{a}(v)$ 
is a good approximation around the outer boundary,
but so small that the approximation \eqref{C-sol} given below is good.
(This range of $N$ exists because the second condition
only requires $N \ll a^2/\ell_p^2$.)
For a given value of $u$,
the $v$-coordinate of the outer boundary of the near-horizon region
will be denoted by $v_{out}(u)$.
It should be the inverse function of $u_{out}(v)$:
$v_{out}(u_{out}(v)) = v$.

In the conventional model of black holes,
the horizon is assumed to be ``uneventful''
\cite{Davies:1976ei,Fulling:1977jm,Christensen:1977jc,Parentani:1994ij,Frolov:1998wf}.
This means that the vacuum energy-momentum tensor
is not larger than $\mathcal{O}(1/a^4)$
for freely falling observers comoving with the collapsing matter.
After the coordinate transformation to the light-cone coordinates $(u, v)$, 
the conditions for uneventful horizons are given by
\cite{Fulling:1977jm,Christensen:1977jc}
\begin{align}
\langle T_{uu}\rangle &\sim \mathcal{O}(C^2/a^4), 
\label{Tuu}
\\
\langle T_{uv}\rangle &\sim \mathcal{O}(C/a^4),
\label{Tuv}
\\
\langle T_{vv}\rangle &\sim \mathcal{O}(1/a^4),
\label{Tvv}
\\
\langle T_{\th\th}\rangle
&
\sim \mathcal{O}(1/a^2).
\label{Tthth}
\end{align}
This can be computed either
by solving the geodesic equation for freely falling observers,
or by computing the transformation factor $dU/du$ between the coordinate $u$
and the light-cone coordinate $U$ suitable for the comoving observers
(see eq.\eqref{dUdu}).

The component $\langle T_{uu} \rangle$ \eqref{Tuu}
is nearly vanishing around the horizon because $C \ll 1$ there,
otherwise there would be a huge outgoing energy flux
for observers comoving with the collapsing matter.\footnote{
Using eq.\eqref{dUdu} below, 
we obtain $\langle T_{UU} \rangle \simeq C^{-2} \langle T_{uu}\rangle$,
where $U$ is the light-cone coordinate suitable for freely falling observers.
$\langle T_{UU} \rangle$ would become very large for $C\ll1$
unless $\langle T_{uu}\rangle \propto C^2$ as in eq.\eqref{Tuu}.}
On the other hand, 
in the large distance limit $r \rightarrow \infty$ where $C\to1$,
$\langle T_{uu} \rangle$ approaches $\mathcal{O}(1/a^4)>0$,
corresponding to Hawking radiation at large distances,
while $\langle T_{uv} \rangle \sim \langle T_{vv} \rangle \sim \langle T_{\th\th}\rangle \sim 0$
in the asymptotically flat region,
so the energy of the system must decrease.
This means that the ingoing energy flux $\langle T_{vv} \rangle$
must be negative around the horizon for energy conservation.
This negative ingoing energy is also the necessary condition for the appearance of 
a time-like trapping horizon (see e.g. Ref.\cite{Ho:2019kte}). 
The outer boundary of the near-horizon region,
which stays outside the trapping horizon,
is also time-like. 
Hence,
any point $(u, v)$ inside the trapping horizon satisfies
\begin{align}
v < v_{ah}(u) < v_{out}(u), 
\qquad
u > u_{ah}(v) > u_{out}(v),
\label{timelikeness}
\end{align}
where $v_{ah}(u)$ and $u_{ah}(v)$ are the $v$ and $u$ coordinates
of the trapping horizon at given $u$ or $v$, respectively.
(See Fig.\eqref{UV}.)

In this paper,
we will only consider the range of near-horizon region in which 
\begin{align}\label{uuvvrange}
u - u_{out}(v) \ll \mathcal{O}(a^3/\ell_p^2),
\qquad
v_{out}(u) - v \ll \mathcal{O}(a^3/\ell_p^2).
\end{align}
For our conclusion about the breakdown of the effective field theory,
we will only need the knowledge of the spacetime geometry
in a much smaller neighborhood.

The energy-momentum tensor \eqref{Tuu}--\eqref{Tthth} for an uneventful horizon
corresponds to the Unruh vacuum
and is often viewed as an implication of the equivalence principle.
However, 
we will see in Sec.\ref{effectivefieldtheory} that an uneventful horizon
always evolves into an eventful horizon at a later time
for a generic effective theory soon after the collapsing matter enters the near-horizon region.

\subsection{Solution of $C(u, v)$}
In this subsection,
we review the solution of $C(u, v)$ in the metric \eqref{metric}
\cite{HMY-1,ShortDistance}.
Two of the semi-classical Einstein equations 
$G_{uv} = \kappa \langle T_{uv} \rangle$ and $G_{\th\th} = \kappa \langle T_{\th\th} \rangle$
can be linearly superposed as \cite{ShortDistance}
\be
\del_u\del_v \Sigma(u, v) = \frac{C(u, v)}{4r^2(u, v)}
+ \frac{\kappa C(u, v)}{8} \left(\langle T^{\mu}{}_{\mu} \rangle 
- 6 \langle T^{\th}{}_{\th} \rangle\right),
\label{Sigma-eq}
\ee
where $\Sigma$ is defined by
\be
C(u, v) \equiv \frac{e^{\Sigma(u, v)}}{r(u, v)}.
\label{C-sigma}
\ee
For the Schwarzschild solution \eqref{C-Schwarzschild}--\eqref{Sch-drdu},
$\Sigma(u, v)$ becomes
\begin{align}
\Sigma = \log(r-a) \simeq \frac{v - u}{2a} - 1 + \log(a)
\label{Sigma-Schwarzschild}
\end{align}
in the near-horizon region.

We shall carry out our perturbative calculation
in the double expansion of $\ell_p^2/a^2$ and $C(u, v)$.
The red-shift factor $C(u, v)$ is of $\mathcal{O}(\ell_p^2/a^2)$
around the trapping horizon,
but $C(u, v)$ gets exponentially smaller as one goes deeper into the near-horizon region.
(See eqs.\eqref{C-sigma}, \eqref{Sigma-Schwarzschild} above and eq.\eqref{C-sol} below.) 
With more focus on the deeper part of the near-horizon region,
every quantity is first expanded in powers of $C(u, v)$,
and then the coefficients of each term in powers of $\ell_p^2/a^2$.

We expand $\Sigma$ as
\begin{align}
\Sigma = \Sigma_0 + \Sigma_1 + \Sigma_2 + \cdots,
\label{Sigma}
\end{align}
where $\Sigma_0 \sim \mathcal{O}(C^0)$,
$\Sigma_1 \sim \mathcal{O}(C)$, $\Sigma_2 \sim \mathcal{O}(C^2)$, etc.
At the leading order,
eq.\eqref{Sigma-eq} indicates
\begin{align}
\del_u\del_v \Sigma_0 &= 0,
\label{Sigma0-eq}
\end{align}
where we have used eqs.\eqref{Tuv}, \eqref{Tthth} to estimate
$\langle T^{\mu}{}_{\mu} \rangle$ and $\langle T^{\th}{}_{\th} \rangle$.

Eq.\eqref{Sigma0-eq} can be easily solved by $\Sigma_0 = B(u) + \bar{B}(v)$
for two arbitrary functions $B(u)$ and $\bar{B}(v)$.
Without loss of generality, 
we can define $a(u)$ and $\bar{a}(v)$ by
\begin{align}\label{def_aa}
a(u) = -\frac{1}{2B'(u)},
\qquad 
\bar{a}(v) = \frac{1}{2\bar{B}'(v)},
\end{align}
so that
\begin{align}
\Sigma_0(u, v) &= \Sigma_0(u_{\ast}, v_{\ast}) 
- \int_{u_{\ast}}^u \frac{du'}{2a(u')} - \int_v^{v_{\ast}} \frac{dv'}{2\bar{a}(v')}.
\label{sigma0-sol}
\end{align}
Comparing eq.\eqref{sigma0-sol} with the Schwarzschild case \eqref{Sigma-Schwarzschild},
we can see that $a(u)$ and $\bar{a}(v)$ should be interpreted as generalizations
of the notion of Schwarzschild radius for the dynamical solution.
Roughly speaking,
one may interpret $a(u)$ as the Schwarzschild radius observed
at the outer boundary of the near-horizon region along
an infinitesimal slice from $u$ to $u+du$,
and $\bar{a}(v)$ the Schwarzschild radius observed
at the outer boundary along an infinitesimal slice
from $v$ to $v+dv$.
(As the Schwarzschild metric is static,
the Schwarzschild radius can be determined
on a single slice of the spacetime.
But in the dynamical case,
choosing a fixed $u$ or a fixed $v$ gives different geometries
and thus different Schwarzschild radii.)
See Ref.\cite{ShortDistance} for more discussion.
At the leading order,
$\bar{a}(v)$ agrees with the mass parameter 
in the special case of the ingoing Vaidya metric (see App.\ref{IVM}). 
In the classical limit $\ell_p^2/a^2 \rightarrow 0$,
both $a(u)$ and $\bar{a}(v)$ approach to
the Schwarzschild radius $a$.

More precisely,
since $\del_u\Sigma_0$ is independent of $v$,
it can be identified with $\del_u\Sigma$
at the outer boundary of the near-horizon region
(where the Schwarzschild solution is a good approximation).
Similarly,
$\del_v\Sigma$ is independent of $u$ and it can also be determined this way.
We can think of $a(u)$ and $\bar{a}(v)$ as the Schwarzschild radii for the best fit
of the Schwarzschild metric on constant-$u$ and constant-$v$ slices
in a small neighborhood around the boundary of the near-horizon region.
For a larger $N$ (see eq.\eqref{outer-boundary-condition}),
the Schwarzschild approximation is better at the outer boundary of the near-horizon region,
hence there should be a smaller difference between $a(u_{out}(v))$ and $\bar{a}(v)$. 
In App.\ref{a-abar}, we derive the relation
\begin{align}
\frac{a(u_{out}(v))}{\bar{a}(v)} \simeq 1 + \mathcal{O}\left(\frac{1}{N}\right)
\label{a=abar}
\end{align}
between $a(u)$ and $\bar{a}(v)$ at the boundary of the near-horizon region.
The functional forms of $a(u)$ and $\bar{a}(v)$ are determined by 
differential equations \eqref{dr0dv}, \eqref{dadu} to be derived below.

It is then deduced from eqs.\eqref{C-sigma}, \eqref{Sigma}, and \eqref{sigma0-sol}
that the solution of $C(u, v)$
can be approximated by \cite{ShortDistance}
\begin{align}
C(u, v) &\simeq C_{\ast} \frac{r_{\ast}}{r(u, v)}
\exp\left[- \int^u_{u_{\ast}} \frac{du'}{2a(u')} - \int_v^{v_{\ast}} \frac{dv'}{2\bar{a}(v')}\right]
\left[1 + \mathcal{O}(C)\right],
\label{C-sol}
\end{align}
where $C_{\ast} \equiv C(u_{\ast}, v_{\ast})$
and $r_{\ast} \equiv r(u_{\ast}, v_{\ast})$
for an arbitrary reference point $(u_{\ast}, v_{\ast})$ in the near-horizon region.
For given $u$,
since $v < v_{out}(u)$ inside the near-horizon region \eqref{timelikeness},
eq.\eqref{C-sol} implies that $C(u, v) < C(u, v_{out}(u))$,
where $C(u, v_{out}(u))$ can be estimated by
the Schwarzschild approximation \eqref{C-Schwarzschild}
to be $\sim N\ell_p^2/\bar{a}^2$,
using eq.\eqref{outer-boundary-condition}.
Due to the exponential form of $C(u, v)$ \eqref{C-sol},
the value of $C$ is exponentially smaller as we move
deeper inside the near-horizon region,
i.e. for larger $u - u_{\ast}$ or larger $v_{\ast} - v$.


\subsection{Solution of $r(u, v)$}
The solution of $r(u, v)$ in the metric \eqref{metric}
can be readily derived using the solution of $C(u, v)$ \eqref{C-sol}. 
We start by estimating the orders of magnitude of $\del_u r$ and $\del_v r$.
From the definition of the Einstein tensor $G_{uu}$ for the metric \eqref{metric}:
\be\label{Guu}
G_{uu} \equiv \frac{2\del_u C \del_u r}{Cr} - \frac{2\del_u^2 r}{r},
\ee
the semi-classical Einstein equation
$G_{uu} = \kappa \langle T_{uu} \rangle$
and eq.\eqref{Tuu},
we derive
\begin{align}
\del_u\left(\frac{\del_u r}{C}\right) &= -\frac{r}{2C} G_{uu} = - \frac{\kappa r}{2C} \langle T_{uu}\rangle
\sim \mathcal{O}(\ell_p^2 C/a^3),
\end{align}
which can be integrated as
\begin{align}
\del_u r(u, v) &= - \frac{\kappa}{2} C(u, v) \int_{u_{\ast}}^u du' \, \frac{r(u', v)}{C(u', v)}T_{uu}(u', v)
+ \frac{C(u, v)}{C(u_{\ast}, v)}\del_u r(u_{\ast}, v).
\label{delur}
\end{align}
In this expression,
we can choose $u_{\ast} = u_{out}(v)$ so that $(u_{\ast}, v)$ is located
on the outer boundary of the near-horizon region.
The values of $C(u_{\ast}, v_{\ast})$ and $\del_u r$ can thus 
be estimated in the Schwarzschild approximation
according to eqs.\eqref{C-Schwarzschild} and \eqref{Sch-drdu},
so the 2nd term in eq.\eqref{delur} is of $\mathcal{O}(C(u, v))$.
One can use $C(u, v)$ \eqref{C-sol} to check that the first term in eq.\eqref{delur} 
is much smaller than the 2nd term.
(See App.\ref{integral-evaluation}.)
Thus, we find
\begin{align}
\del_u r(u, v)
\sim \mathcal{O}(C(u, v)).
\label{delur-order}
\end{align}

In a similar manner as App.\ref{integral-evaluation}, 
we can use $C(u, v)$ \eqref{C-sol} again to derive 
from $G_{vv} = \kappa \langle T_{vv} \rangle$ and eq.\eqref{Tvv} that
\begin{align}
\del_v r(u, v) &= \frac{\kappa}{2} C(u, v) \int^{v_{\ast}}_v dv' \, \frac{r(u, v')}{C(u, v')}T_{vv}(u, v')
\lesssim \mathcal{O}(\ell_p^2/a^2),
\label{delvr}
\end{align}
where we chose $v_{\ast} = v_{ah}(u)$ so that
the reference point $(u, v_{ah}(u))$ is located on the trapping horizon,
and used the condition $\del_v r(u, v_{ah}(u)) = 0$ on the trapping horizon.



As the linear combination \eqref{Sigma-eq} of 
the semi-classical Einstein equations $G_{uv} = \kappa \langle T_{uv} \rangle$
and $G_{\th\th} = \kappa \langle T_{\th\th} \rangle$
is already satisfied by $C(u, v)$ \eqref{C-sol},
only one more independent linear combination of them is needed.
We choose to look at 
\begin{align}
G_{uv} \equiv
\frac{C}{2r^2} + \frac{2\del_u r\del_v r}{r^2} + \frac{2\del_u\del_v r}{r} 
= \kappa \langle T_{uv} \rangle.
\end{align}
Using eqs.\eqref{Tuv}, \eqref{C-sol}, \eqref{delur-order}, and \eqref{delvr}
to estimate the order of magnitude
of each term in this equation,
we find it to be dominated by the two terms $C/2r^2$ and $2\del_u\del_v r/r$,
so that
\begin{align}
\del_u\del_v r(u, v)
&\simeq - \frac{C(u, v)}{4r(u, v)} + \mathcal{O}(\ell_p^2 C/a^3).
\label{ddr=C}
\end{align}
To integrate this, we suppose that the black hole evaporates 
in the time scale of $\Delta u, \Delta v \sim \mathcal{O}(a^3/\ell^2_p)$
as usual \cite{Hawking-Radiation}. 
Hence, the $u$ and $v$ derivatives of $a(u)$ and $\bar{a}(v)$
introduce additional factors of $\mathcal{O}(\ell_p^2/a^3)$ 
because the two radii are approximately the Schwarzschild radius. 
Also, from eqs.\eqref{delur-order} and \eqref{delvr}, 
the $u$ and $v$ derivatives of $r(u, v)$ lead to extra factors of $\mathcal{O}(\ell_p^2/a^3)$. 
On the other hand, with $C(u, v)$ given by eq.\eqref{C-sol},
its $u$ and $v$ derivatives produce only factors of $-1/2a$ and $1/2\bar{a}$, respectively. 
Thus, the functions $a(u)$, $\bar{a}(v)$ and $r(u, v)$
are approximately constant in comparison with $C(u, v)$, 
and eq.\eqref{ddr=C} can be solved by
\begin{align}
\del_u r(u, v) &\simeq
- \frac{\bar{a}(v)}{2r(u, v)}C(u, v) + f_1(u) + \mathcal{O}(\ell_p^2 C/a^2),
\label{delur-1}
\\
\del_v r(u, v) &\simeq
\frac{a(u)}{2r(u, v)}C(u, v) + f_2(v) + \mathcal{O}(\ell_p^2 C/a^2)
\end{align}
for arbitrary functions $f_1(u)$ and $f_2(v)$.
However,
comparing the first equation \eqref{delur-1} with eq.\eqref{delur-order},
we see that $f_1(u)$ has to vanish,
because a function of $u$ cannot go to $0$ as fast as $C(u, v)$
in the limit $(v_{\ast} - v) \rightarrow \infty$.
According to eq.\eqref{delvr},
we find $f_2(v) \lesssim \mathcal{O}(\ell_p^2/a^2)$.

The consistent solution to the two equations above is
\begin{align}
r(u, v)
&\simeq r_0(v) + \frac{a(u)\bar{a}(v)}{r_0(v)}C(u, v) 
+ \mathcal{O}\left(\frac{\ell_p^2}{a} C\right),
\label{r-sol0}
\end{align}
where the function $r_0(v)$ can be determined as follows. 
First, in the classical limit,
$C = 1 - a/r$ \eqref{C-Schwarzschild} can be rewritten as 
$r = a + rC \simeq a + aC$ near $r\sim a$,
which resembles eq.\eqref{r-sol0}.
Since both $a(u)$ and $\bar{a}(v)$
coincide with the Schwarzschild radius $a$ in the classical limit,
we have $r_0 = a$ in the limit as well. 
Therefore, turning on quantum effect,
we expect $r_0(v)$ to be approximately equal to $\bar{a}(v)$.
To estimate the order of magnitude of the difference $r_0(v) - \bar{a}(v)$,
we plug the solution $r(u, v)$ \eqref{r-sol0} into the condition \eqref{outer-boundary-condition}
on the outer boundary of the near-horizon region for $u = u_{out}(v)$.
Then we find the relation
\be
r_0(v) - \bar{a}(v) = \left(1-\frac{a(u_{out}(v))\bar{a}(v)}{r_0^2(v)}\right)\frac{N\ell_p^2}{\bar{a}(v)}
\lesssim \mathcal{O}\left(\frac{N\ell_p^2}{\bar{a}(v)}\right),
\label{r0-sol}
\ee
where we used eqs.\eqref{C-Schwarzschild}, \eqref{outer-boundary-condition}
to evaluate $C(u_{out}(v), v)$.
Using eq.\eqref{r0-sol} in eq.\eqref{r-sol0},
we find
\begin{align}
r(u, v) &\simeq \bar{a}(v) + a(u) C(u, v)
+ \mathcal{O}\left(\frac{N\ell_p^2}{a} C\right)
\label{r-sol}
\\
&\simeq \bar{a}(v) + \mathcal{O}\left(\frac{N\ell_p^2}{\bar{a}(v)}\right)
\label{r-sol-simpler}
\end{align}
in the near-horizon region.


Let us now determine the time-evolution of the functions $a(u)$ and $\bar{a}(v)$.
Plugging eqs.\eqref{C-sol} and \eqref{r-sol} back into the semi-classical Einstein equations
$G_{uu} = \kappa \langle T_{uu} \rangle$ (with eq.\eqref{Guu}), 
we can check that this equation is trivially satisfied at the leading order 
in the $\ell_p^2/a^2$ expansion and does not impose any constraint on $a(u)$. 
Similarly, we can see that $G_{vv} = \kappa \langle T_{vv} \rangle$ gives
\begin{align}\label{GTvv}
\frac{\bar{a}'(v)}{\bar{a}^2(v)} - \frac{2\bar{a}''(v)}{\bar{a}(v)}
\simeq \kappa \langle T_{vv}(u, v) \rangle.
\end{align}
As the left-hand side of this equation is $u$-independent,
$\langle T_{vv}(u, v) \rangle$ is $u$-independent at the leading order
in the near-horizon region. 
Recall the uneventful condition \eqref{Tvv} 
that $\langle T_{vv}(u, v) \rangle$ must be negative and of $\mathcal{O}(1/a^4)$. 
It can be expressed as
\be
\langle T_{vv}(u, v)\rangle \simeq  - \frac{\sigma}{\kappa \bar{a}^4(v)}
\ee
for some parameter $\sigma \sim \mathcal{O}(1)$. 

Now, we consider an adiabatic process \cite{Barcelo:2010xk} 
of Hawking radiation for which $|\bar{a}'/\bar{a}| \gg |\bar{a}''|$.
Eq.\eqref{GTvv} then becomes
\be
\frac{d\bar{a}(v)}{dv} \simeq \kappa \bar{a}^2(v) \langle T_{vv} \rangle
\simeq - \frac{\sigma\ell_p^2}{\bar{a}^2(v)},
\label{dr0dv}
\ee
which determines the functional form of $\bar{a}(v)$.
The function $a(u)$ is approximately equal to $\bar{a}(v)$ at $u = u_{out}(v)$
due to eq.\eqref{a=abar},
so
\begin{align}
\frac{d\bar{a}(v)}{dv} &\simeq
\frac{da(u_{out}(v))}{dv} 
= \frac{du_{out}(v)}{dv} \left.\frac{da(u)}{du}\right|_{u=u_{out}(v)}
\simeq \left.\frac{da(u)}{du}\right|_{u=u_{out}(v)},
\end{align}
where we used eq.\eqref{dudv-out}.
Using eq.\eqref{a=abar} on the right-hand side of eq.\eqref{dr0dv},
we find
\begin{align}
\frac{da(u)}{du} \simeq
- 
\frac{\sigma\ell_p^2}{a^2(u)}.
\label{dadu}
\end{align}

\subsection{Near-horizon geometry}
\label{metricunderhorizon}

The solution for $C$ \eqref{C-sol} can now be further simplified
using the solution for $r$ \eqref{r-sol} as
\begin{align}
C(u, v) &\simeq C(u_{\ast}, v_{\ast}) 
\frac{\bar{a}(v_{\ast})}{\bar{a}(v)}
\exp\left[- \int^u_{u_{\ast}} \frac{du'}{2a(u')} - \int_v^{v_{\ast}} \frac{dv'}{2\bar{a}(v')}\right]
\left[1 + \mathcal{O}(C)\right].
\label{C-sol-again}
\end{align}
This and the solution of $r(u, v)$ given by eq.\eqref{r-sol} define the metric \eqref{metric} for
the geometry of the near-horizon region,
with $\bar{a}(v)$, $a(u)$ satisfying eqs.\eqref{dr0dv}, \eqref{dadu}.

In the following,
we will also need the Christoffel symbol of the metric \eqref{metric}:
\begin{align}
\Gamma_{uu}^u &= \frac{\del_u C(u, v)}{C(u, v)} 
= - \frac{1}{2a(u)}\left[1 + \mathcal{O}(C)\right],
\label{Gammauuu}
\\
\Gamma_{vv}^v &= \frac{\del_v C(u, v)}{C(u, v)} 
= \frac{1}{2\bar{a}(v)}\left[1 + \mathcal{O}\left(\frac{\ell_p^2}{a^2}\right)\right],
\label{Gammavvv}
\end{align}
with other components
$\Gamma_{uv}^u, \Gamma_{uv}^v, \Gamma_{uu}^v, \Gamma_{vv}^u$
vanishing.

Finally, note that the characteristic length scale for all curvature invariants is still $\bar{a}$,
e.g.
\begin{align}
R \simeq - \frac{2}{\bar{a}^2},
\qquad
R_{\mu\nu}R^{\mu\nu} \simeq \frac{16}{\bar{a}^4},
\qquad
R_{\mu\nu\lam\rho}R^{\mu\nu\lam\rho} \simeq \frac{8}{\bar{a}^4}.
\label{curvature}
\end{align}
As we will see below, nevertheless, 
the metric \eqref{r-sol} and \eqref{C-sol-again} together with 
the quantum effect of non-renormalizable operators lead to a non-trivial physical effect.

\section{Breakdown of effective theory}
\label{effectivefieldtheory}

For the low-energy effective theory of, say, a 4D massless scalar field $\phi$,
we have an action
\begin{align}
S &= \int d^4 x \sqrt{-g} \, {\cal L},
\label{S}
\end{align}
with a Lagrangian density given as
a $1/M_p$-expansion:
\begin{align}
{\cal L} &= 
\frac{1}{2}g^{\mu\nu}\nabla_{\mu}\phi \nabla_{\nu}\phi
+ \frac{a_1}{4!} \phi^4 + a_2 R \phi^2 
+ \frac{1}{M_p^2}\left[
b_1 (\nabla^2\phi)(\nabla^2\phi)
+ b_2 \phi^6 + b_3 (\nabla^2 R) \phi^2
+ \cdots \right]
\nn \\
&\quad
+ \frac{1}{M_p^4}\left[
c_1 g^{\mu\nu}(\nabla_{\mu}\nabla^2\phi)(\nabla_{\nu}\nabla^2\phi)
+ c_2 \phi^8
+ c_3 (\nabla^2 R) (\nabla\phi)^2
+ \cdots \right]
+ \cdots.
\label{action-4D}
\end{align}
(Assuming the symmetry $\phi \rightarrow -\phi$,
we omit terms of odd powers of $\phi$ for simplicity.)
The dimensionless parameters $a_1, a_2, b_1, b_2, \cdots$
are the coupling constants in a perturbation theory.
Higher-dimensional terms are suppressed by higher powers of $1/M_p$.

For a given physical state,
it is normally assumed that
all higher-dimensional (non-renormalizable) interactions,
which are suppressed by powers of $1/M_p$,
only have negligible contributions to its time evolution.
We will show below that,
since the effective-field-theoretic derivation of Hawking radiation
involves high-frequency modes of quantum fluctuations,
there are in fact higher-dimensional operators 
in the effective Lagrangian \eqref{S}
that contribute to large probability amplitudes of particle creation 
from the Unruh vacuum in the near-horizon region. 
We will see that this particle creation
makes the uneventful horizon ``eventful'' or even ``dramatic''. 

\subsection{Free-field quantization in the near-horizon region}\label{free-field}
In this subsection,
we introduce the quantum-field-theoretic formulation
for the computation of the amplitudes mentioned above.
It is essentially the same as the standard formulation
for the derivation of Hawking radiation (see e.g. Ref.\cite{Brout:1995rd}). 
The difference is that we shall consider 
the background geometry given in Sec.\ref{geometry},
instead of the static Schwarzschild background.

For a massless scalar field $\phi$
in the near-horizon region,
we shall focus on its fluctuation modes with spherical symmetry.
It is convenient to define
\be
\varphi(u, v) \equiv r(u, v) \phi(u, v)
\ee
for the $s$-wave modes. 
For the metric \eqref{metric},
the free-field equation $\nabla^2\phi = 0$ is equivalent to
\begin{align}
\del_u\del_v \varphi - \frac{\del_u\del_v r}{r} \varphi = 0.
\label{free-field-eq}
\end{align}
According to eqs.\eqref{ddr=C}, \eqref{r0-sol} and \eqref{r-sol},
it becomes
\begin{align}
\del_u\del_v \varphi + \frac{C(u, v)}{4\bar{a}^2}\varphi \simeq 0
\end{align}
in the near-horizon region.
The free-field equation is thus well approximated by
\be
\del_u\del_v \varphi \simeq 0
\ee
deep inside the near-horizon region where $C$ is exponentially small. 
Therefore, the general solution there is given by
\be
\varphi \simeq \int_0^{\infty} \frac{d\om}{2\pi} \,
\frac{1}{\sqrt{2\om}}\left(
e^{-i\om U(u)}a_{\om} + e^{i\om U(u)}a^{\dag}_{\om} 
+ e^{-i\om V(v)}\tilde{a}_{\om} + e^{i\om V(v)}\tilde{a}^{\dag}_{\om}
\right).
\label{phi_dec}
\ee
Here, $U(u)$ and $V(v)$ are arbitrary functions of $u$ and $v$, respectively.
The creation and annihilation operators $\{a_{\om}, a^{\dag}_{\om}\}$
and $\{\tilde a_{\om}, \tilde a^{\dag}_{\om}\}$
satisfy 
\begin{equation}
[a_{\om_1}, a^{\dag}_{\om_2}] = 2\pi \d(\om_1 - \om_2),~~
[\tilde a_{\om_1}, \tilde a^{\dag}_{\om_2}] = 2\pi \d(\om_1 - \om_2),
\label{aa-comm}
\end{equation}
with the rest of the commutators vanishing.

In principle, we can use any functions $U(u)$ and $V(v)$
as the outgoing and ingoing light-cone coordinates.
We shall choose the light-cone coordinates $U$ and $V$
so that the vacuum $|0\rangle$ defined by
\begin{align}
a_{\om} | 0 \rangle =\tilde a_{\om} | 0 \rangle = 0 
\qquad \forall \om \geq 0
\label{def-0}
\end{align}
is the Minkowski vacuum of the infinite past
before the gravitational collapse starts.
This is the vacuum which evolves into Hawking radiation at large distances
after it falls in from the past infinity, passes the origin, and then moves out
\cite{Hawking-Radiation}.
We assume that this vacuum $|0\rangle$ is the quantum state of the near-horizon region. 
It is equivalent to the Unruh vacuum
--- the vacuum state for freely falling observers
at an uneventful horizon \cite{Unruh:1977ga}.

The relation between the coordinates $U$ and $u$
can be derived easily by considering the special case 
when the collapsing matter is a spherical thin shell at the speed of light,
and identifying $U$ with the retarded light-cone coordinate
of the flat Minkowski spacetime inside the collapsing shell
\cite{Davies:1976ei,Unruh:1976db}
as follows.
\footnote{
If the collapsing shell is not thin,
it only introduces negligible corrections to
the relation between $U$ and $u$
in the near-horizon region.} 
The trajectory of the areal radius $R_s(u) = r(u, v_s)$ of the thin shell
(where $v_s$ is the $v$-coordinate of the thin shell)
satisfies 
\begin{align}
\frac{dR_s}{dU} = - \frac{1}{2},
\end{align}
where we used $r(U, V) = (V-U)/2$ in the flat space.
It also satisfies
\begin{align}
\frac{dR_s}{du} = \del_u r(u, v_s) 
\simeq - \frac{1}{2} C(u, v_s),
\end{align}
following eqs.\eqref{r-sol}, \eqref{dadu}, and \eqref{C-sol-again}.
The two equations above imply
\begin{align}
\frac{dU(u)}{du} \simeq C(u, v_s),
\label{dUdu}
\end{align}
and hence the conditions \eqref{Tuu}--\eqref{Tthth}
simply mean that
$T_{UU} \sim T_{UV} \sim T_{VV} \sim T_{\th\th} \sim \mathcal{O}(1/a^4)$. 

We decompose the field $\phi = \varphi/r$ \eqref{phi_dec} into 
the outgoing and ingoing modes.
In the near-horizon region, 
the outgoing modes can be expanded in two bases: 
\begin{align}
\phi_{out}(u, v) &= \int_0^{\infty} \frac{d\om}{2\pi} \frac{1}{\sqrt{2\om}}
\frac{1}{r(u, v)} \left(
e^{-i\om U(u)}a_{\om} + e^{i\om U(u)}a^{\dag}_{\om} \right)
\label{phi-U}
\\
&= \int_0^{\infty} \frac{d\om}{2\pi} \frac{1}{\sqrt{2\om}}
\frac{1}{r(u, v)} \left(
e^{-i\om u}c_{\om} + e^{i\om u}c^{\dag}_{\om} \right).
\label{phi-out}
\end{align}
The two expressions above are related by the coordinate transformation \eqref{dUdu} 
and the creation and annihilation operators
$\{c_{\om}, c^{\dag}_{\om}\}$ satisfy
\begin{equation}
[c_{\om_1}, c^{\dag}_{\om_2}] = 2\pi \d(\om_1 - \om_2),
\qquad
[c_{\om_1}, c_{\om_2}] = [c^{\dag}_{\om_1}, c^{\dag}_{\om_2}] = 0.
\label{cc}
\end{equation}
They are related to $\{a_{\om}, a^{\dag}_{\om}\}$ via a Bogoliubov transformation
\begin{align}
c_{\om} &= 
\int_0^{\infty} d\om' \, \left(A_{\om\om'} a_{\om'} + B_{\om\om'} a^{\dag}_{\om'}\right),
\label{Bogo1}
\\
c^{\dag}_{\om} &= 
\int_0^{\infty} d\om' \, \left(A^{\ast}_{\om\om'} a^{\dag}_{\om'} + B^{\ast}_{\om\om'} a_{\om'}\right).
\label{Bogo2}
\end{align}
The equality between eqs.\eqref{phi-U} and \eqref{phi-out} determines
the coefficients $A_{\om\om'}$ and $B_{\om\om'}$ as
\begin{align}
A_{\om\om'} = \frac{1}{2\pi} \sqrt{\frac{\om}{\om'}} \int_{-\infty}^{\infty} du \, e^{i\om u-i\om' U(u)},
\qquad
B_{\om\om'} = \frac{1}{2\pi} \sqrt{\frac{\om}{\om'}} \int_{-\infty}^{\infty} du \, e^{i\om u+i\om' U(u)}.
\end{align}

For the vacuum state $|0\rangle$ defined by eq.\eqref{def-0},
it is natural to define a 1-particle state
\begin{align}
|\om\rangle_a \equiv \sqrt{2\om} a^{\dag}_{\om}|0\rangle.
\label{om-a}
\end{align}
On the other hand,
we also consider the 1-particle state
\begin{align}
|\om\rangle^c &\equiv
{\cal N} \, \sqrt{2\om} \, c_{\om} | 0 \rangle
= {\cal N} \, \int_0^{\infty}
d\om' \, \sqrt{\frac{\om}{\om'}} \, B_{\om\om'} | \om' \rangle_a,
\label{1-particle-state}
\end{align}
which is a superposition of the 1-particle states $|\om'\rangle_a$.

In the calculation below, 
we will need to evaluate the quantity
${}^c\langle \om | \phi | 0 \rangle$,
and hence we have to estimate the matrix $B_{\om\om'}$
appearing in eq.\eqref{1-particle-state}.
As we will see, 
only a short time scale $\Delta u \sim \mathcal{O}(a \log a/\ell_p)$ is relevant
to our calculation below.
(See eq.\eqref{u-v-cond}.)
Within this time scale,
the black-hole mass does not change much
so that $a(u)$ remains roughly the same value
that we will simply denote by $a$.
Therefore, from eq.\eqref{dUdu} and eqs.\eqref{dadu}--\eqref{C-sol-again},
we have approximately
\be
U(u) \simeq U_h - c_0 e^{- \frac{u}{2a}}
\label{U-u}
\ee 
for an arbitrary constant $U_h$,
and $c_0$ is determined by eq.\eqref{dUdu} to be
\begin{align}
c_0 
= 2a C(u_{\ast}, v_s) e^{\frac{u_{\ast}}{2a}}.
\label{c0}
\end{align}
The Bogoliubov coefficients can be approximated by
\footnote{See, for example, Ref.\cite{Brout:1995rd}.}
\begin{align}
A_{\om\om'} &\simeq
\frac{a}{\pi} \sqrt{\frac{\om}{\om'}} e^{-i\om' U_h} \left(\frac{1}{\om' c_0}\right)^{-i2a\om}
e^{\pi a\om} \Gamma(-i2a\om),
\label{A-approx}
\\
B_{\om\om'} &\simeq
e^{2i\om' U_h} e^{-2\pi a\om} A_{\om\om'}.
\label{B-approx}
\end{align}
One then deduces from eqs.\eqref{A-approx} and \eqref{B-approx} that
\begin{align}
\int_0^{\infty} d\om'' \, A_{\om\om''}A^{\ast}_{\om'\om''} &\simeq
\frac{\d(\om - \om')}{1-e^{-4\pi a\om}},
\label{AAast}
\\
\int_0^{\infty} d\om'' \, A_{\om\om''} B_{\om'\om''} &\simeq 0.
\label{AB}
\end{align}

The normalization factor ${\cal N}$ defined in eq.\eqref{1-particle-state}
is fixed by the condition
\begin{align}
{}^c\langle \om|\om' \rangle^c = 4\pi \om \delta(\om - \om')
\label{omc-norm}
\end{align}
(following eqs.\eqref{1-particle-state}, \eqref{B-approx},
\eqref{AAast} and \eqref{omc-norm})
to be
\begin{align}
{\cal N} =
\sqrt{e^{4\pi a\om} - 1}.
\label{N}
\end{align}
Then we find
\begin{align}
{}^c\langle \om | \phi | 0 \rangle
\hide{
&= {\cal N} \int \frac{d\om'}{2\pi} \sqrt{\frac{\om}{\om'}} \, \frac{1}{r}
\left(
{}_a \langle 0 | c^{\dag}_{\om}c^{\dag}_{\om'} | 0 \rangle e^{i\om' u}
+ {}_a \langle 0 | c^{\dag}_{\om}c_{\om'} | 0 \rangle e^{-i\om' u}
\right)
\nn \\
&= \sqrt{e^{4\pi a\om}-1}
\int d\om' d\om'' \sqrt{\frac{\om}{\om'}} \, \frac{1}{r}
\left(
B^{\ast}_{\om\om''} A^{\ast}_{\om'\om''} e^{i\om' u}
+ B^{\ast}_{\om\om''} B_{\om'\om''} e^{-i\om' u}
\right)
\nn \\
&\simeq \sqrt{e^{4\pi a\om}-1}
\int d\om' \sqrt{\frac{\om}{\om'}} \, \frac{1}{r}
\left(
\frac{\d(\om - \om')}{e^{4\pi a\om}-1} e^{i\om' u}
\right)
\nn \\
&\simeq 
\frac{1}{\sqrt{e^{4\pi a\om}-1}}
\frac{1}{r} e^{i\om u}
\nn \\
}
&\simeq 
\frac{1}{{\cal N} r} \,
e^{- i\om u}.
\label{om-phi-0}
\end{align}
\hide{
using the relation
\begin{align}
\frac{1}{\sqrt{\om}} \, e^{i\om U} = 
\int_0^{\infty} d\om' \; \frac{1}{\sqrt{\om'}} \,
\left(A^{\ast}_{\om'\om} e^{i\om' u} + B_{\om'\om} e^{-i\om' u}\right).
\end{align}
}

In eq.\eqref{def-0},
we have introduced the $(U, V)$ coordinates
as the light-cone coordinates used to define
the Minkowski vacuum of the infinite past $|0\rangle$.
Therefore, it is natural to identify the $V$-coordinate
in the same approximation scheme as
\begin{align}
V \simeq V_h + 2a e^{\frac{v - v_s}{2a}},
\label{V-v}
\end{align}
so that we have
\begin{align}
C \simeq \frac{dU}{du} \frac{dV}{dv},
\label{C-UV}
\end{align}
which leads to
\begin{align}
ds^2_{(2D)} \simeq - dU dV.
\label{ds2-UV0}
\end{align}
This means that the $(U, V)$ coordinates are those
of a freely falling observer who describes the spacetime
locally as flat.
Note that the $(U, V)$ coordinates take essentially the same form
as the usual Kruskal coordinates.
They play the role of the Kruskal coordinates in the dynamical spacetime.

For the ingoing modes, we have
\begin{align}
\phi_{in}(u, v) &= \int_0^{\infty} \frac{d\om'}{2\pi} \frac{1}{\sqrt{2\om'}}
\frac{1}{r(u, v)} \left(
e^{-i\om' V(v)}\tilde{a}_{\om'} + e^{i\om' V(v)}\tilde{a}^{\dag}_{\om'} \right),
\label{phi-in}
\end{align}
and there are counterparts of the equations shown above for the outgoing modes. 
In particular, we can define the 1-particle states
\begin{align}
|\om\rangle_{\tilde{a}} \equiv \sqrt{2\om} \tilde{a}^{\dag}_{\om}|0\rangle.
\label{1-particle-in}
\end{align}
But we will not need the operators $\tilde{c}_{\om}$, $\tilde{c}^{\dag}_{\om}$
defined with respect to the light-cone coordinates $(u, v)$
for the ingoing modes.

\subsection{Transition amplitude}
\label{highenergyevents}

In general, the effective Lagrangian \eqref{action-4D}
includes all local invariants.
As examples,
we consider a class of higher-dimensional,
higher-derivative local observables of dimension $[M]^{2n+k+l+1}$:
\begin{align}
\hat{O}_{\{m\}l} &\equiv
g^{\mu_1\nu_1}\cdots g^{\mu_n\nu_n}
\left(\nabla_{\mu_1}\cdots\nabla_{\mu_n} \phi_1\right)
\left(\nabla_{\nu_1}\cdots\nabla_{\nu_{m_1}} \phi_2\right)
\left(\nabla_{\nu_{m_1+1}}\cdots\nabla_{\nu_{m_1+m_2}} \phi_2\right)
\cdots
\nn \\
& \qquad \cdots
\left(\nabla_{\nu_{n-m_k+1}}\cdots\nabla_{\nu_n} \phi_2\right)
\phi_3^l
\qquad \qquad \qquad (k, m_1, \cdots, m_k \geq 1; \; l \geq 0),
\label{On}
\end{align}
where $n \equiv \sum_{i=1}^{k} m_i$.
The fields $\phi_1$, $\phi_2$, and $\phi_3$ are all massless scalars,
and all equations for $\phi$ in Sec.\ref{free-field} 
apply to $\phi_1$, $\phi_2$ and $\phi_3$.
(The calculation below will be essentially the same if $\phi_1 = \phi_2 = \phi_3$.) 

Due to the dynamical background,
this operator \eqref{On} introduces
a time-dependent perturbation to the free field theory.
The corresponding interaction term in the action \eqref{S} is
\begin{align}
\frac{\lam_{\{m\}l}}{M_p^{2n+k+l-3}} 
\int d^4 x \sqrt{-g}
\, \hat{O}_{\{m\}l},
\label{interaction-op}
\end{align}
where $\lam_{\{m\}l}$ is a coupling constant of $\mathcal{O}(1)$.
We shall consider its matrix element
\begin{align}
{\cal M}_{\{m\}l} \equiv 
\frac{\lam_{\{m\}l}}{M_p^{2n+k+l-3}}
\int_{\cal V} dx^4 \sqrt{-g} \, \langle f | \hat{O}_{\{m\}l} | i \rangle
\label{Mn-0}
\end{align}
integrated over a spacetime region ${\cal V}$,
where $|i\rangle$ is the Unruh vacuum
and $|f\rangle$ is a multi-particle state to be defined below.

For ${\cal V} = (t_0, t_1)\times $space
($t$ is a time coordinate),
the matrix element \eqref{Mn-0} can be interpreted as 
the transition amplitude from the initial state $|i\rangle$ at $t = t_0$
to the final state $|f\rangle$ at $t = t_1$
in the first-order time-dependent perturbation theory.
We will show below that
${\cal M}_{\{m\}l}$ becomes exponentially large
when the collapsing matter enters deeply inside the near-horizon region.

One might naively think that such a transition amplitude must be small
since the initial state $|i\rangle$ is the Unruh vacuum.
As the typical length scale is $\mathcal{O}(a)$
for the small curvature \eqref{curvature}, 
one expects that ${\cal M}_{\{m\}l}$ is
$\sim \mathcal{O}((\ell_p/a)^{2n+k+l-3})$
by dimensional analysis. 
However, it turns out that ${\cal M}_{\{m\}l} $ becomes large
as a joint effect of the peculiar geometry in the near-horizon region
and the quantum fluctuation of the matter field. 

The Hilbert space of the perturbative quantum field theory
is the tensor product of the Fock spaces of the 3 fields $\phi_1$, $\phi_2$ and $\phi_3$.
The initial state is the tensor product of the Unruh vacuum for each field,
\begin{align}
| i \rangle
&\equiv 
| 0 \rangle \otimes | 0 \rangle \otimes | 0 \rangle.
\label{i}
\end{align}
The final state of interest is of the form
\begin{align}
| f \rangle
&\equiv 
| \om \rangle^c \otimes | \om'_1, \cdots, \om'_k \rangle_{\tilde{a}} \otimes
| \om_1, \om_2, \cdots, \om_l \rangle_{\tilde{a}}.
\label{f}
\end{align}
Here, $|\om\rangle^c$ is the superposition \eqref{1-particle-state}
of outgoing modes of $\phi_1$,
$|\om'_1, \cdots, \om'_k \rangle_{\tilde{a}}$ the $k$-particle state
as a generalization of the 1-particle state \eqref{1-particle-in}
for the ingoing modes of $\phi_2$,
and $|\om_1, \om_2, \cdots, \om_l \rangle_{\tilde{a}}$
the $l$-particle state of the ingoing modes of $\phi_3$,
respectively.

We shall choose 
\begin{align}
\om \sim \mathcal{O}(1/a)
\label{om-others}
\end{align}
for the state $|\om\rangle^c$.
Notice that the prediction of the spectrum of Hawking radiation relies
on a field-theoretic calculation of
$\langle 0 | c^{\dag}_{\om}c_{\om'} | 0 \rangle
= {}^c\langle \om|\om'\rangle^c/(2\sqrt{\om\om'}{\cal N}^2)$.
If the state $|\om\rangle^c$ is not well-defined in the low-energy effective theory
at least for $\om \sim \mathcal{O}(1/a)$,
our understanding of Hawking radiation would be reduced to almost nothing.
This state $|\om\rangle^c$ must be allowed in the effective theory; 
otherwise, the existence of Hawking radiation would be dubious. 

On the other hand,
the values of $\om'_1, \cdots, \om'_k$
and $\om_1, \om_2, \cdots, \om_l$
will not play an important role
in showing the matrix element \eqref{Mn-0} to be large.
We shall simply choose
\begin{align}
\om'_1 \simeq \om'_2 \simeq \cdots \simeq \om'_k \simeq
\om_1 \simeq \om_2 \simeq \cdots \simeq \om_l \simeq 0
\label{small-om}
\end{align}
for simplicity.


Due to the $s$-wave reduction,
all the spacetime indices $\mu_i, \nu_i$ are either $u$ or $v$,
and each factor of $g^{uv}$ contributes a factor of $C^{-1}(u, v)$.
The covariant derivatives $\nabla_u$, $\nabla_v$
involve derivatives $\del_u$, $\del_v$,
which contribute factors of frequencies
$\om, \om'_1, \cdots, \om'_k, \om_1, \cdots, \om_l$.
Hence,
the transition amplitude is the integral of a polynomial in $\om, \om'_1, \cdots, \om'_k, \om_1, \cdots, \om_l$,
apart from an overall factor including$e^{-i\om u} e^{i\sum_{a = 1}^k \om'_a v} e^{i\sum_{i=1}^l\om_i v}$.
To show that the transition amplitude \eqref{Mn-0} is large,
it is sufficient to focus on a term with given powers of $\om, \om'_1, \cdots, \om'_k, \om_1, \cdots, \om_l$,
as they are independent free parameters.
We shall focus on the terms with the largest power of $\om$
but independent of $\om'_1, \cdots, \om'_k, \om_1, \cdots, \om_l$.
It is
\begin{align}
{\cal M}_{\{m\}l} &\sim \frac{\lam_{\{m\}l}\ell_p^{2n+k+l-3}}{{\cal N}} \, \om^n
\int_{\cal V} du dv \,
\frac{1}{C^{n-1}} \,
\frac{1}{r^{l-1}} \,
\left[\prod_{i=1}^{k}\left(\nabla_v^{m_i}\frac{1}{r}\right)\right]
e^{- i\om u}.
\label{Mn-01}
\end{align}

The expression \eqref{Mn-01} for the transition amplitude ${\cal M}_{\{m\}l}$ tells us that,
the integral over ${\cal V}$ is dominated by the contribution of the region
where $\del_V r$ is large and $C$ is small.
On the other hand,
it is unclear why a small conformal factor $C$,
which has no particular local meaning for a freely falling observer,
leads to a large transition amplitude.
To understand the reason why the transition amplitude is large
from the viewpoint of freely falling observers,
we will rewrite this expression \eqref{Mn-01} in the next subsection
in terms of the coordinates $(U, V)$ suitable for freely falling observers.

\subsection{Comments on the amplitude ${\cal M}_{\{m\}l}$}

We study here the properties of the amplitude \eqref{Mn-01}
and explain the strategy of its evaluation for the next subsection.

\subsubsection{Amplitudes in static background}
\label{static-back}

Before we estimate ${\cal M}_{\{m\}l}$ \eqref{Mn-01}
for the dynamical background, 
we check that it vanishes for any static background,
including the Schwarzschild metric.
Let $t$ be the time coordinate with translation symmetry,
the functions $C$, $r$ and the Christoffel symbol are all independent of $t$. 
The only $t$-dependence in ${\cal M}_{\{m\}l}$ 
is thus the exponential factor $e^{-i\om u} \propto e^{-i\om t}$, 
so we have
\begin{align}
{\cal M}_{\{m\}l} \propto \int^{t_1}_{t_0} dt' \, e^{-i\om t'} = 
\frac{e^{-i\om t_1} - e^{-i\om t_0}}{-i\om}.
\label{energy-conserv}
\end{align}
for ${\cal V} = (t_0, t_1)\times$ space.
The transition amplitude is non-zero
as an artifact of the boundaries at $t_0$ and $t_1$.
It vanishes, for instance,
if $\om$ is quantized to satisfy the periodic boundary condition.
Hence,
unless the time-dependence of the dynamical background is taken into account,
the transition amplitude ${\cal M}_{\{m\}l}$ vanishes
for suitable boundary conditions.
Note that if ${\cal M}_{\{m\}l} \neq 0$,
it means that particles (and hence their energies) are created
out of the vacuum in ${\cal V}$.
Therefore,
${\cal M}_{\{m\}l} = 0$ is simply a consequence of energy conservation
in the region with time-translation symmetry.

\subsubsection{Large amplitudes in dynamical background}

When the back-reaction of Hawking radiation is included,
the factor 
\begin{align}
\frac{1}{C^{n-1}} \,
\frac{1}{r^{l-1}} \,
\left[\prod_{i=1}^{k}\left(\nabla_v^{m_i}\frac{1}{r}\right)\right]
\label{quantity}
\end{align}
in the transition amplitude \eqref{Mn-01} has no time-translation symmetry,
so its integral with the phase $e^{-i\om u}$ is in general non-zero.

Naively,
even though ${\cal M}_{\{m\}l}$ \eqref{Mn-01} is no longer exactly $0$,
one might still expect that it is negligible due to the overall factor $\ell_p^{2n+k+l-3}$.
However,
this factor can be compensated by $C^{-n+1}$ in eq.\eqref{quantity},
since the conformal factor $C$ can be arbitrarily close to $0$
deep inside the near-horizon region.
According to the solution of $C$ \eqref{C-sol-again},
a displacement of $u$ or $v$ by
a small amount $2k a\log(a/\ell_p)/(n-1)$
is enough to compensate a factor of $(\ell_p/a)^k$.

\hide{
We emphasize here that the back-reaction of the vacuum energy momentum tensor
to the background geometry discussed in Sec.\ref{geometry} is crucial.
If we use the Schwarzschild solution as an approximation,
we have $C = 1 - a/r$ and the factor \eqref{quantity} becomes
\begin{align}
\frac{1}{C^{n-1}} \,
\frac{1}{r^{l-1}} \,
\left[\prod_{i=1}^{k}\left(\nabla_v^{m_i}\frac{1}{r}\right)\right]
&=
\frac{1}{C^{n-1}} \,
\frac{1}{r^{l-1}} \,
\left[
\prod_{i=1}^{k}\left(
\frac{(-1)^{m_i}m_i !}{2^{m_i}r^{m_i + 1}} \, C^{m_i}
\right)
\right]
\nn \\
&=
\frac{(-1)^{n} \prod_{i=1}^{k} m_i !}{2^{n}r^{n + k + l - 1}} \, C
\propto C,
\label{quantity-naive}
\end{align}
where we have used the relation
\begin{align}
\nabla_v^{m_i}\frac{1}{r} =
\frac{(-1)^{m_i}m_i !}{2^{m_i}r^{m_i + 1}} \, \left(1 - \frac{a}{r}\right)^{m_i}
\label{dv1/r}
\end{align}
for the Schwarzschild solution.
Eq.\eqref{quantity-naive} implies that the matrix element
is small when the conformal factor $C$ is small.
The overall factor of $C^{-(n-1)}$ in the matrix element \eqref{Mn-01}
is cancelled by the factors $\nabla_v^{m_i}\frac{1}{r} \propto C^{-m_i}$.
This calculation based on the Schwarzschild solution explains why
people could not have found large matrix elements due to
higher-derivative interactions in the past
before the analytic result in Sec.\ref{geometry} was available.
We will see below that,
using the result of Sec.\ref{geometry},
eq.\eqref{dv1/r} would be modified
so that $\nabla_v^{m_i}\frac{1}{r}$ is suppressed by powers of $1/M_p$
but not by $C^{m_i}$.
The matrix element becomes large as a result of the overall factor $C^{-(n-1)}$.
}

The claim that the matrix element can become large due to a small $C$ is unsettling
because the appearance of the arbitrarily small conformal factor $C$
relies on the choice of the $(u, v)$ coordinate system.
If we use the Kruskal coordinate $(U, V)$
(given by eqs.\eqref{U-u} and \eqref{V-v}),
the metric becomes locally eq.\eqref{ds2-UV0};
the conformal factor is $1$.
A natural question is then:
How can the amplitude become large?
As the operator $\hat{\cal O}_{\{m\}l}$ \eqref{On} is by definition a scalar,
its integral over a given region of spacetime
is independent of the choice of coordinates.
To understand the physics better,
let us first answer this question
by analyzing M in terms of $(U,V)$.

For the locally flat metric \eqref{ds2-UV0},
we have $\nabla_U = \del_U$ and $\nabla_V = \del_V$.
When we rewrite the amplitude \eqref{Mn-01} 
in terms of the $(U, V)$ coordinate system,
it becomes
\begin{align}
{\cal M}_{\{m\}l} 
&\sim
\frac{2\pi \lam_{\{m\}l}}{M_p^{2n+k+l-3}} 
\int_{\cal V} dU dV  \,
r^2 (g^{UV})^n \langle f | (\partial_U^n\phi_1)
\left[\prod_{i=1}^{k}(\partial_V^{m_i}\phi_2)\right]
\phi_3^l
| i \rangle
\nn \\
&\sim
\frac{2\pi (-2 i)^n {\cal N} \lam_{\{m\}l}}{M_p^{2n+k+l-3}}
\int_{\cal V} dU dV  \,
\int_0^{\infty} d\om_U \;
\sqrt{\frac{\om}{\om_U}} B_{\om\om_U}^{\ast}
\frac{\om_U^n}{r^{l-1}}
\left[\prod_{i=1}^{k}\partial_V^{m_i}\frac{1}{r}\right]
e^{i\om_U U}
\nn \\
&\simeq
\alpha_{\{m\}l} \,
\int_0^{\infty} d\om_U \,
\om_U^{n-1-i2a\om} \,
{\cal A}_{\{m\}l}(\omega_U),
\label{UV-integral}
\end{align}
where
\begin{align}
\alpha_{\{m\}l} &\equiv
2(-2i)^n a\om {\cal N} \lam_{\{m\}l} \Gamma(i2a\om)\,
e^{-\pi a\om} c_0^{- i2a\om} \ell_p^{2n+k+l-3},
\label{alphaml}
\\
{\cal A}_{\{m\}l}(\omega_U) &\equiv
\int_{\cal V} dU dV  \,
\frac{1}{r^{l-1}}
\left[\prod_{i=1}^{k}\partial_V^{m_i}\frac{1}{r}\right]
\, e^{i\om_U (U-U_h)}.
\label{Aml}
\end{align}
To derive this expression,
we have used eqs.\eqref{i}, \eqref{f} for the states $|i\rangle$, $|f\rangle$,
eq.\eqref{On} for the operator $\hat{\cal O}_{\{m\}l}$,
and eqs.\eqref{aa-comm}, \eqref{phi-U}, \eqref{om-a}, \eqref{1-particle-state},
\eqref{A-approx}, \eqref{B-approx}, and \eqref{small-om}
to evaluate the matrix element.
This is simply eq.\eqref{Mn-01} written in terms of the Kruskal coordinates.

Indeed the expression \eqref{UV-integral}
does not explicitly involve any exponentially growing factor.
To see how the factor $C^{-(n-1)}$ in eq.\eqref{Mn-01}
is hidden in the expression above,
we should carry out the integration over $\om_U$.
The $\om_U$-integral of the form
\begin{align}
\int_0^{\infty} d\om_U \,
\om_U^{m-i2a\om} e^{i\om_U (U-U_h)}
\end{align}
in eq.\eqref{UV-integral} (with $m = n - 1$)
can be evaluated using the saddle point approximation.
The saddle point is
\begin{align}
\om_{Us} 
= - \left(\om + \frac{im}{2a}\right)\left(\frac{dU}{du}\right)^{-1},
\label{saddle}
\end{align}
where we have used eq.\eqref{U-u}.
It is important to note that $\om_{Us}$ is large
when the blue-shift factor $(dU/du)^{-1}$ is large.

The integral over $\om_U$ in eq.\eqref{UV-integral} 
is thus approximately
\begin{align}
\int_0^{\infty} d\om_U \;
\om_U^{m-i2a\om} e^{i\om_U (U-U_h)}
\sim
\om_{Us}^{m+1-i2a\om} e^{i\om_{Us} (U-U_h)}
= 
\left[
- \left(\om + \frac{im}{2a}\right)\left(\frac{dU}{du}\right)^{-1}
\right]^{m+1-i2a\om} e^{-m+i2a\om},
\label{int-omU}
\end{align}
up to a factor of $\mathcal{O}(1)$.

On the other hand, 
the factor $(dV/dv)^{-n+1}$ appears from
\begin{align}
dV \left[\prod_{i=1}^{k}\partial_V^{m_i}\frac{1}{r}\right]
= dv \left(\frac{dV}{dv}\right)^{-(n-1)}
\left[\prod_{i=1}^{k}\nabla_v^{m_i}\frac{1}{r}\right]
\label{int-omV}
\end{align}
in eq.\eqref{UV-integral},
where $n \equiv \sum_{i = 1}^k m_i$.
Thus, 
$dU/du$ in eq.\eqref{int-omU} (for $m = n-1$)
and $dV/dv$ in eq.\eqref{int-omV}
produce the hidden factor $C^{-(n-1)}$
according to eq.\eqref{C-UV}.
This explains how the large factor $C^{-(n-1)}$
arises in the calculation in terms of the $(U, V)$-coordinates.
\footnote{
There may be other factors of $C$ to a positive power
in the calculation of the amplitude,
but we will see that,
generically,
with a sufficiently large order of derivatives,
the amplitude involves a negative power of $C$.
}

Strictly speaking,
the region ${\cal V}$ of integration needs to be infinitely large
so that the Fourier transform with respect to $\omega_U$ is well defined.
For a finite ${\cal V}$,
we should use a suitable complete basis of functions in ${\cal V}$.
A simple example is when ${\cal V}$ is a rectangular region
with periodic boundary conditions such that 
$e^{i\om_U U}$ can be used as the basis
but $\omega_U$ is discretized.
(See, e.g. eq.\eqref{quant-omU}.)
In this case,
we should first integrate over $(U, V)$ to find ${\cal A}_{\{m\}\ell}(\om_U)$ \eqref{Aml},
assuming that $\om_U$ is properly discretized,
and then replace $\int_0^{\infty} d\om_U$ in eq.\eqref{UV-integral}
by a sum $\sum_{\om_U}$ over discretized values of $\om_U$.
For a sufficiently large region ${\cal V}$,
the sum over $\sum_{\om_U}$ should be well approximated by the integral,
so we expect that the conclusion above for infinite ${\cal V}$
remains qualitatively correct for a finite ${\cal V}$.
In App.\ref{example-calc},
we consider the discretization of $\omega$ for a finite region
and carry out the explicit calculation of the transition amplitude
to demonstrate the general expectation described above.

\subsection{Example: thin shell and ${\cal M}_{\{1\cdots 1\}0}$}
\label{example}

To demonstrate explicitly that the magnitude of the amplitude \eqref{UV-integral} 
becomes large as the collapsing matter falls further inside the near-horizon region
(so that the conformal factor $C$ becomes small),
we study a simple example here.
We consider a thin shell collapsing at the speed of light
along the curve $V = V_s$
and investigate a special class of higher-derivative interactions
\begin{align}
\hat{\cal O}_{\{1\cdots 1\}0} \equiv g^{\mu_1\nu_1}\cdots g^{\mu_n\nu_n}
\left(\nabla_{\mu_1}\cdots\nabla_{\mu_n}\phi_1\right)
\left(\nabla_{\nu_1}\phi_2\right)\cdots\left(\nabla_{\nu_n}\phi_2\right).
\label{O10}
\end{align}
This is the case of $\hat{\cal O}_{\{m\}l}$ \eqref{On}
with $m_i = 1$ for $i = 1, \cdots, k = n$ and $l = 0$,
and we assume $n > 2$. 

\begin{figure}
\center
\includegraphics[scale=0.5,bb=0 0 500 250]{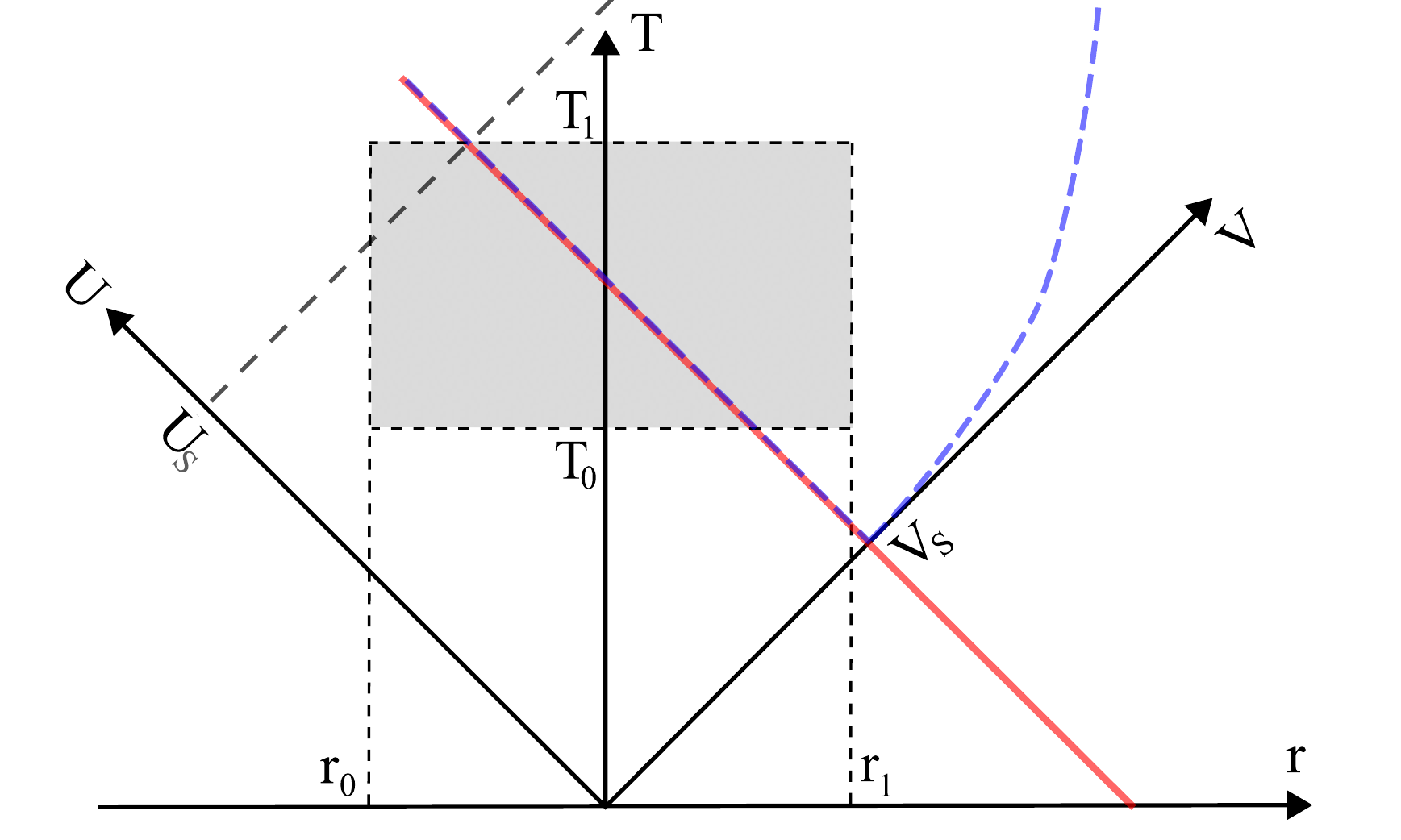}
\caption{\small
The shaded rectangle is the region ${\cal V}$ defining the transition amplitude.
The blue dash curve for $V > V_s$ is the outer trapping horizon,
and the inner trapping horizon coincides with
the collapsing thin shell at the speed of light (the red line at $V = V_s$).
The contribution of this domain ${\cal V}$ to the matrix element
is dominated by a neighborhood of the point $(U_s, V_s)$.
}
\label{UV-1}
\vskip1em
\end{figure}

In terms of the time coordinate $T$ defined by
\be
T \equiv \frac{U + V}{2},
\label{T-def}
\ee
we choose ${\cal V}$ to be a rectangle $(T_0, T_1)\times (r_0, r_1)$
which covers a large space.
(See Fig.\ref{UV-1}.)
It can be divided into the following 4 parts:
(i) the space outside the near-horizon region ($V > V_s$ and $r \gg a$),
(ii) the near-horizon region ($V > V_s$ and $r \sim a$),
(iii) the thin shell ($V = V_s$),
and (iv) the flat space inside the shell ($V < V_s$).

In App.\ref{example-calc},
we evaluate the order of magnitude of ${\cal M}_{\{1\cdots 1\}0}$ as
\begin{align}
{\cal M}_{\{1\cdots 1\}0}
\sim
\frac{2\pi (2n-1) (n-3)! \lam_{\{1\cdots 1\}0} e^{-\pi a\om} \zeta(-(n-3-i2a\om))
\ell_p^{3n-3}}{2^{n-2} \sinh(2\pi a\om) a^{3n-3}} \,
\frac{(T_1 - T_0)}{a} \;
C^{-(n-2)}(u_s, v_s),
\label{M110}
\end{align}
up to a factor of $\mathcal{O}(1)$.
The dominant contribution comes from the region (iv),
more specifically the corner with the maximal value $u_s$ of $u$ and minimal value $v_s$ of $v$
along the trajectory of the collapsing shell in ${\cal V}$.
(Recall eq.\eqref{timelikeness} and see Fig.\ref{UV-1}.)
This implies that the shape of the region ${\cal V}$ is not important.
\footnote{
It was pointed out in Ref.\cite{Giddings:2006sj} that
a large matrix element is obtained
(for an operator without higher derivatives)
when only the space outside the event horizon is integrated over,
but it is merely an artifact of the boundary condition at the event horizon,
and this large contribution is cancelled by
the space inside the event horizon.
Here we take ${\cal V}$ to cover the four different regions (i) -- (iv)
to rule out the possibility that the matrix element becomes large
due to an artificial boundary condition.
}

We note that the conformal factor $C(u_s, v_s)$
in ${\cal M}_{\{1\cdots 1\}0}$
scales by a factor of $e^{- \Delta/2a}$
under a shift in $u_s$ by $\Delta$.
It implies that the amplitude \eqref{M110} is exponentially larger
when the collapsing shell is deeper inside the near-horizon region.

$C^{-1}(T_1-T_0)$ is the time duration
of the region ${\cal V}$ for a distant observer,
and we will be interested in a duration of time of the order of the scrambling time,
$\Delta t \sim a \log(a/\ell_p)$ \cite{Sekino:2008he}.
Here we assume that
\be
C^{-1}(T_1 - T_0)/a \gtrsim\mathcal{O}(1).
\label{TT}
\ee
Hence, 
for $\om \sim 1/a$ and $n > 2$ but not too large
\footnote{
For large $n$,
the amplitude is further enhanced by other factors in eq.\eqref{M110}.
},
the transition amplitude \eqref{M110} can be estimated as
\begin{align}
{\cal M}_{\{1\cdots 1\}0}
&\sim 
\frac{\ell_p^{3n-3}}{a^{3n-3}} \,
\frac{C^{-1}(u_s, v_s)(T_1 - T_0)}{a} \,
C_{\ast}^{-(n-3)} \, e^{(n-3)\frac{u_s - u_{\ast} + v_{\ast} - v_s}{2a}}
\nn \\
&\gtrsim 
\frac{\ell_p^{n+3}}{a^{n+3}} \,
e^{(n-3)\frac{u_s - u_{\ast} + v_{\ast} - v_s}{2a}},
\label{M110-1}
\end{align}
Here, 
we have chosen the reference point $(u_{\ast}, v_{\ast})$
to be located on the trapping horizon so that
\begin{align}
C_{\ast} \equiv C(u_{\ast}, v_{\ast})
\sim \mathcal{O}\left(\frac{\ell_p^2}{a^2}\right),
\label{Cast}
\end{align}
which comes from eqs.\eqref{C-Schwarzschild} and \eqref{outer-boundary-condition}.
Therefore,
the matrix element ${\cal M}_{\{1\cdots 1\}0}$ is larger than $\mathcal{O}(1)$
when
\begin{align}
u_s-u_{\ast} + v_{\ast} - v_s \geq 2 \left(\frac{n+3}{n-3}\right) a\log(a/\ell_p).
\label{u-v-cond}
\end{align}
For example,
let us take the reference point to be the point 
where the shell crosses the trapping horizon
(the trapping horizon emerges at this moment $u = u_{\ast}$)
so that $v_{\ast} = v_s$.
Then,
the matrix element becomes larger than $\mathcal{O}(1)$
after an elapse of time $\Delta u \equiv u_s - u_{\ast}$ of the same order of magnitude
as the scrambling time $\mathcal{O}(a\log(a/\ell_p))$.
(This is consistent with the range of the near-horizon region \eqref{timelikeness}.)


A large matrix element ${\cal M}_{\{1\cdots 1\}0}$ implies a large
transition amplitude from $|i\rangle$ \eqref{i} to $|f\rangle$ \eqref{f}.
As the thin shell falls further deep under the apparent horizon,
the energy flux of the created outgoing particles in $|f\rangle$ grows exponentially.
This can be identified with the firewall \cite{firewall,firewall-B}
because the saddle-point frequency $\om_{Us}$ is trans-Planckian
with respect to comoving observers.
(There will be more discussion on this in the next subsection.)
According to eq.\eqref{u-v-cond},
it should appear within the scrambling time
after the shell enters the apparent horizon.
In fact,
we will see below that the transition amplitudes become large
for many other higher-derivative interactions
even before the condition \eqref{u-v-cond} is met.

Finally, 
we discuss the contribution of the collapsing matter to the amplitude ${\cal M}_{\{m\}l}$.
In the case above,
the shell matter has no contribution to eq.\eqref{M110}
because $\delta (V-V_s)$ does not appear (see App.D).
In a generic matter configuration,
a higher energy density can induce a larger transition amplitude.
On the other hand,
even if the prefactor in eq.\eqref{M110-1} is much smaller
(say, by a factor of $\ell_p^{2(n-1)}/a^{2(n-1)}$
as it would be if only the contribution of the region outside the collapsing shell is included,
see App.D),
the amplitude still becomes large within the same order of magnitude as the scrambling time.
Furthermore,
the cancellation between the matter contribution and the vacuum one generically
dose not occur because it requires a fine tuning
(see App.D for more discussion).
Therefore, the conclusion about the scrambling time should be robust
independently of the matter configuration.
In the following,
we will consider only the contribution of the near-horizon region
outside the collapsing matter for simplicity.

\subsection{Firewall}

Now we consider another class of operators different from the example above.
We show that the matrix element becomes huge at the moment
when the collapsing matter enters the near-horizon region,
and this corresponds to the firewall.

Consider the operators
\begin{align}
\hat{\cal O}_{\{n/2,n/2\}0} \equiv g^{\mu_1\nu_1}\cdots g^{\mu_n\nu_n}
\left(\nabla_{\mu_1}\cdots\nabla_{\mu_n}\phi_1\right)
\left(\nabla_{\nu_1}\cdots\nabla_{\nu_{n/2}}\phi_2\right)
\left(\nabla_{\nu_{n/2+1}}\cdots\nabla_{\nu_n}\phi_2\right),
\end{align}
which is $\hat{\cal O}_{\{m\}l}$
with $k = 2$, $m_1 = m_2 = n/2$ and $l = 0$.
($n > 2$ and $n$ is even.)
The corresponding matrix element \eqref{UV-integral} is given by
eqs.\eqref{alphaml} and \eqref{Aml} as
\begin{align}
{\cal M}_{\{n/2,n/2\}0} \sim 
\ell_p^{2n-1} \int d\om_U \, \om_U^{n-1-i2a\om}
\int dU dV \,
r \left(\partial_V^{n/2}\frac{1}{r}\right)^2
e^{i\om_U(U - U_h)}.
\label{366}
\end{align}

As we commented at the end of Sec.\ref{example},
when higher-derivatives of the quantum fields are involved,
the contribution of the matter to the matrix element 
depends on the details of the matter configuration.
To avoid this uncertainty,
in this section,
we will focus on the contribution of the near-horizon region,
even though the contribution of the region occupied by the collapsing matter
can be larger.

Using the equation
\begin{align}
\del_{V}^m\frac{1}{r} \simeq
\left(\frac{dV}{dv}\right)^{-m}
\frac{(m-1)!}{(-2\bar{a})^{m-1}}\frac{\sigma\ell_p^2}{\bar{a}^2 r^2},
\label{delVn}
\end{align}
derived from eqs.\eqref{r-sol}, \eqref{dr0dv} and \eqref{V-v},
we evaluate eq.\eqref{366} as
\begin{align}
{\cal M}_{\{n/2,n/2\}0} \sim 
\frac{\ell_p^7}{a^7} \left(\frac{C^{-1}(u_s, v_s)(T_1 - T_0)}{a}\right)
\, e^{(n-3)\frac{u_s - u_{\ast} + v_{\ast} - v_s}{2a}}.
\label{367}
\end{align}
The derivation is essentially the same as that of eq.\eqref{M110} in App.\ref{example-calc},
but only with the contribution of the near-horizon region taken into consideration.

For a reasonably long period of time $C^{-1}(T_1 - T_0) \gtrsim a$ \eqref{TT}
for the region ${\cal V}$ from the viewpoint of a distant observer
(which is an extremely short time $(T_1 - T_0)$ for a freely falling observer),
the amplitude ${\cal M}_{\{n/2,n/2\}0}$ is larger than $\mathcal{O}(1)$ as long as
\begin{align}
u_s - u_{\ast} + v_{\ast} - v_s \geq \frac{14}{n-3} \, a \log(a/\ell_p).
\label{u-v-cond-2}
\end{align}
This is a smaller lower bound than eq.\eqref{u-v-cond} for $n > 4$
but still the same order of magnitude as the scrambling time for finite $n$.

\hide{
Interestingly,
in the large $n$ limit, 
the right hand side of eq.\eqref{u-v-cond-2} goes to $0$.\footnote{
For a complete discussion of the large $n$ limit,
we also need to take into consideration
the numerical factor $((n/2 - 1)!)^2$
and the coupling constant $\lam_{\{n/2,n/2\}0}$,
which were ignored in eq.\eqref{367}.
}
This implies that the matrix element can be large already
as soon as the collapsing matter falls into the near-horizon region.
}

The final state $|f\rangle$ for the exponentially increasing transition amplitude
includes the outgoing mode $|\om\rangle^c$ \eqref{1-particle-state},
which is a superposition of 1-particle states $|\om_U\rangle_a$
for freely moving observers.
For comoving observers,
the dominant frequency $\om_U$ of these 1-particle states
is the saddle point \eqref{saddle} with the magnitude
\begin{align}
|\om_U| \sim |\om| \left(\frac{dU}{du}\right)^{-1}
\sim \frac{a}{\ell_p^2} \, e^{(u - u_{\ast} + v_{\ast} - v_s)/2a},
\end{align}
which is trans-Planckian at $u = u_s$
well before eq.\eqref{u-v-cond-2} is satisfied.
Hence the large matrix elements imply the presence of a firewall
as a flux of trans-Planckian particles in the comoving frame.

Before the effective theory breaks down,
there are particle creations with exponentially increasing probability,
although the prediction of a firewall as a Planckian energy flux
is not reliable.
Depending on the UV-theory
(or some of the coupling constants $\lam_{\{m\}l}$ at large $n$),
the energy flux of the created particles may or may not become Planckian
before the effective theory breaks down.
It is possible that
the UV theory admits a new effective theory
that will become appropriate to describe what happens afterwards.

\subsection{Viewpoint of freely falling observers}
\label{trans-Planckian}

The saddle point approximation \eqref{saddle} shows that 
the matrix element ${\cal M}_{\{m\}l}$ is dominated by
contributions of trans-Planckian modes $|\om_U\rangle_a$.
The physical reason behind this is clear.
The Hawking radiation is dominated by modes
with frequencies $\om \sim \mathcal{O}(1/a)$ at large distances. 
Tracing these wave packets backwards to the near-horizon region,
they are blue-shifted to trans-Planckian frequencies $\om_U$.

If the trans-Planckian modes are removed from the effective theory,
the matrix elements would not become large,
but it also implies that
there would be no Hawking radiation either.
This is reminiscent of the trans-Planckian problem 
\cite{trans-Planckian-1}.

Note that
we have chosen to consider the 1-particle state
$c_{\om}|0\rangle$ in the final state $|f\rangle$
because our understanding of the spectrum of Hawking radiation relies on
the quantity $\langle 0|c^{\dag}_{\om}c_{\om}|0\rangle$,
which demands that the state $c_{\om}|0\rangle$ be well-defined.
If the amplitude ${\cal M}_{\{m\}l}$ is considered ill-defined
because of its involvement with the trans-Planckian modes,
the spectrum of Hawking radiation is also ill-defined.
While the derivation of Hawking radiation assumes that
the free-field approximation is good,
the matrix elements ${\cal M}_{\{m\}l}$ can be interpreted as
perturbative corrections to the calculation of the spectrum
$\langle 0|c^{\dag}_{\om}c_{\om}|0\rangle$
of Hawking radiation
by higher-derivative interactions.
Large ${\cal M}_{\{m\}l}$ means that Hawking radiation
is largely corrected.

Therefore, 
assuming Hawking radiation and the uneventful horizon,
we cannot avoid the large matrix elements,
leading to the breakdown of the low-energy effective theory.
On the other hand,
it is possible that, in a self-consistent model,
there is a moderately large energy flux around the horizon
(so that it is not uneventful
but also no trans-Planckian modes)
so that a low-energy effective description is still valid.
\cite{Kawai:2020rmt}.
Alternatively,
another logical possibility is that Hawking radiation stops
while the horizon remains free of the Planckian firewall.
More rigorously,
what we have shown is the incompatibility between
Hawking radiation and uneventful horizon
in the effective-field-theoretic description. 

Incidentally,
as an effort to resolve the trans-Planckian problem,
there have been proposals of alternative derivations of Hawking radiation
which assume non-relativistic dispersion relations 
such that the energy is bounded from above to be cis-Planckian
\cite{trans-Planckian-2}.
They reproduce the same spectrum of Hawking radiation,
but this does not completely resolve
the trans-Planckian problem \cite{trans-Planckian-3}
as the wave numbers can still be arbitrarily large.
In the context of this paper,
it is reasonable to expect that,
since the wave number is still allowed to go to infinity,
there are higher-dimensional operators
(which are no longer required to be Lorentz-invariant)
that produce large transition amplitudes,
and the low-energy effective theory still breaks down.
While this remains to be rigorously proven,
what we have shown is at least that,
for {\em relativistic} low-energy effective theories,
Hawking radiation
(which necessarily includes trans-Planckian modes)
is in conflict with the assumption of an uneventful horizon.

Notice that one should not simply dismiss quantum modes with $\om_U > M_p$
as an attempt to solve the trans-Planckian problem.
There are infinitely many freely falling frames at different velocities.
They are related to one another via a local Lorentz boost
\be
U \rightarrow U' = \sqrt{\frac{1+w}{1-w}} \; U, 
\qquad
V\rightarrow V' = \sqrt{\frac{1-w}{1+w}} \; V
\label{uv-transf}
\ee
for a relative velocity $w$.
A constraint like $\om_U < M_p$ has no locally invariant meaning,
as it can always be violated for any non-zero frequency after a boost.
In contrast,
our calculation is invariant under general coordinate transformations.

The choice of a freely falling frame is related to 
the interpretation of the origin of the large matrix elements.
In our calculations,
the origin of the largeness of the matrix element is 
the largeness of $C^{-1}$.
Equivalently, according to eq.\eqref{C-UV},
it is the largeness of $(dU/du)^{-1}$ in the saddle point \eqref{saddle}
and/or $(dV/dv)^{-1}$ in the derivative $\del_V$.
Which one,
$(dU/du)^{-1}$ or $(dV/dv)^{-1}$,
is large?
The answer depends on the choice of the freely falling frame.
\footnote{
For a freely falling observer comoving with the collapsing matter,
the $v$-coordinate of the observer in this frame is roughly constant.
The $(U, V)$ coordinates suitable for the observer
are given by eqs.\eqref{U-u} and \eqref{V-v},
and the transition amplitude increases with the retarded time $u$ 
mostly due to the increase in $(dU/du)^{-1}$
rather than that in $(dV/dv)^{-1}$.
}
A local Lorentz boost eq\eqref{uv-transf}
changes $(dU/du)^{-1}$ and $(dV/dv)^{-1}$ simultaneously,
making one bigger and the other smaller.

A large $(dU/du)^{-1}$ implies a large dominant frequency $\om_{Us}$ \eqref{saddle}
of the 1-particle states $|\om_U\rangle$ for freely falling observers,
and a large $(dV/dv)^{-1}$ means a large $V$-derivative
of the areal radius $r\simeq r_0(v)$ \eqref{r-sol0},
i.e. a fast deformation of the background geometry.
(The magnitude of $dr/dv$ is as small as $\mathcal{O}(\ell_p^2/a^2)$,
but $dr/dV$ can be larger if $(dV/dv)^{-1}$ is large.)
The collision between
the outgoing quantum fluctuation $|\om_U\rangle_a$
and the ingoing geometric deformation $r_0(v)$ defines
a Lorentz invariant energy scale.
When this Lorentz invariant becomes too large,
the effective theory breaks down.

\section{Conclusion and discussion}  
\label{conclusion}
In this work,
we showed that Hawking radiation is incompatible with the uneventful horizon.
Assuming the validity of the effective-theoretic derivation of Hawking radiation,
the higher-dimensional operators in the effective action change the time evolution of
the Unruh vacuum in the near-horizon region of the dynamical black hole 
so that it evolves into an excited state
with many high-energy particles for freely falling observers. 
The uneventful horizon transitions to an eventful horizon (the firewall),
and ultimately the effective theory breaks down. 

We emphasize that we have only used 
the semi-classical Einstein equation and the conventional formulation
of the quantum field theory for the matter field.
The only novel ingredients are 
(i) the explicit solution of the metric in the near-horizon region
and (ii) the consideration of higher-dimensional operators 
in the effective theory. 

For the first item (i), we used the metric given by eqs.\eqref{C-sol} and \eqref{r-sol}
as a solution to the semi-classical Einstein equation
for the energy-momentum tensor \eqref{Tuu}--\eqref{Tthth} of the uneventful horizon. 
As a result of the negative ingoing energy flux $T_{vv}$, 
the trapping horizon is time-like \cite{Ho:2019kte}, 
with the causal structure of the near-horizon region satisfying eq.\eqref{timelikeness}.
This is crucial for the exponential form of the red-shift factor $C(u, v)$
to lead to the exponentially large transition amplitudes after 
the matter enters the near-horizon region. 

We also emphasize the importance of the dynamical nature of the background geometry.
Had we used the static Schwarzschild solution for the background geometry,
the conformal factor would still have the exponential form,
but the matrix elements would be negligible.

About the item (ii), we considered the quantum effect of the higher-dimensional operators
$\hat{\cal O}_{\{m\}l}$ \eqref{On} for $n > 2$.
These are non-renormalizable operators that 
are normally ignored in the low-energy effective theory 
because they are suppressed by powers of $1/M_p$.
However,
we found that these operators induce large transition amplitudes
related to the creation of particles from the Unruh vacuum, 
in contrast with renormalizable operators. 
A lot of the high-energy particles are created for freely falling observers, 
resulting in the firewall. 
This invalidates the conditions \eqref{Tuu}--\eqref{Tthth} for an uneventful horizon.

Note that no local curvature invariants of the dynamical background are found to be large
in the near-horizon region. 
The high-energy events only arise from the higher-dimensional terms in the effective action,
and their origin is a joint effect of the higher-derivative interactions
and the peculiar geometry of the near-horizon region.

Assuming a persisting Hawking radiation,
together with higher-dimensional operators,
there is a firewall,
and the equivalence principle is violated in the sense that
a freely-falling observer sees particles with high energy.
Indeed,
the equivalence principle is in general violated
by higher-derivative interactions.
This has been shown for classical electromagnetism \cite{Lafrance:1994in}.
Although the equivalence principle is violated,
general covariance, 
including the local Lorentz transformation \eqref{uv-transf},
is preserved.


If there is a firewall,
the trans-Planckian scattering between the firewall 
and the collapsing matter cannot be ignored
because the proper distance between the collapsing matter
and the horizon is of the order of a Planck length \cite{ShortDistance}.
It is possible that, 
through such trans-Planckian scatterings,
the information of the collapsing matter is transferred into the outgoing particles,
and information loss is no longer a necessary consequence 
of black-hole evaporation ---
not until one examines this problem with
a Planck-scale theory such as string theory. 

Another possibility is that we abandon the assumption
of uneventful horizon \eqref{Tuu} -- \eqref{Tthth} from the beginning.
It is then still possible that a consistent low-energy effective theory
describes an evaporating black hole.
A self-consistent scenario is perhaps one that would
have no horizon or trapped region,
such as the model proposed in Ref.\cite{Kawai:2013mda,Kawai:2014afa,Kawai:2020rmt}
(see also \cite{KMY}).
It is also recently argued that a consistent quantum theory of gravity
should always admit the VECRO \cite{Mathur:2020ely},
which will likely modify the conventional energy-momentum tensor.

To conclude,
we have shown that Hawking radiation
and uneventful horizon cannot coexist with each other
over the scrambling time.
The low-energy effective theory breaks down
as a result of time evolution from the Unruh vacuum towards the firewall
due to higher-derivative interactions.
How information is preserved is still a problem,
but it is no longer a paradox.

\section*{Acknowledgement}
We thank Hsin-Chia Cheng, Hsien-chung Kao, Hikaru Kawai, Samir Mathur,
and Yoshinori Matsuo for valuable discussions. 
P.M.H. thanks iTHEMS at RIKEN, Tokyo University, and Kyoto University
for their hospitality during his visits when this project was initiated.
P.M.H.\ is supported in part by the Ministry of Science and Technology, R.O.C. 
and by National Taiwan University. 
Y.Y.\ is partially supported by Japan Society of Promotion of Science (JSPS), 
Grants-in-Aid for Scientific Research (KAKENHI) Grants No.\ 18K13550 and 17H01148. 
Y.Y.\ is also partially supported by RIKEN iTHEMS Program.

\appendix

\section{Ingoing Vaidya metric}
\label{IVM}

We consider the ingoing Vaidya metric
as an example to demonstrate the meanings of 
the generalized Schwarzschild radii $a(u)$ and $\bar{a}(v)$.
The ingoing Vaidya metric
\begin{align}
ds^2 = - \left(1 - \frac{a_0(v)}{r}\right) dv^2 + 2dv dr + r^2 d\Omega^2,
\label{ingoing-Vaidya-metric}
\end{align}
where $a_0(v)$ is proportional to the mass parameter of the black hole,
is a spherically symmetric solution to the Einstein equation
for the energy-momentum tensor
\begin{align}
T_{vv} = \frac{a'_0(v)}{\kappa r^2},
\end{align}
with all other components ($T_{vr}, T_{rr}, T_{\th\th}$ etc.) vanishing.
For $a'_0(v) \sim \mathcal{O}(\ell_p^2/a^2_0)$,
the energy-momentum tensor satisfies the uneventful-horizon condition
\eqref{Tuu} -- \eqref{Tthth},
hence the metric \eqref{ingoing-Vaidya-metric}
is just a special case of the general solution
\eqref{C-sol}, \eqref{r-sol}
in the near-horizon region.

To put the metric \eqref{ingoing-Vaidya-metric} in the form of eq.\eqref{metric},
we plug $r = r(u, v)$ into the metric \eqref{ingoing-Vaidya-metric}
and demand that it agrees with eq.\eqref{metric}.
It is
\begin{align}
2 (\del_u r) du dv + \left[2 (\del_v r) - \left(1 - \frac{a_0(v)}{r}\right)\right] dv^2
= - C du dv,
\end{align}
which means that
\begin{align}
\del_u r &= - \frac{1}{2} C,
\label{delur-IVM}
\\
\del_v r &= \frac{1}{2}\left(1 - \frac{a_0(v)}{r}\right).
\label{delvr-IVM}
\end{align}
It is then easy to check that the solution of $r$ \eqref{r-sol-simpler}
satisfies both conditions above at the leading order 
of the $\kappa$-expansion,
in which $|a'(u)|, |\bar{a}'(v)| \sim \mathcal{O}(\ell_p^2/a^2) \ll 1$,
via the identification
\begin{align}
\bar{a}(v) = a_0(v).
\end{align}
Therefore,
$\bar{a}(v)$ can be identified with the Schwarzschild radius $a_0(v)$
of the ingoing Vaidya metric at the leading order.

On the other hand,
the parameter $a(u)$ is not directly fixed by the ingoing Vaidya metric
because the form of the metric \eqref{metric} is invariant under
a coordinate transformation $u \rightarrow u' = u'(u)$.
The $u$-coordinate in the solution \eqref{C-sol}, \eqref{r-sol-simpler}
has been chosen such that,
on the outer-boundary of the near-horizon region,
it agrees with the $u$ coordinate used in the Schwarzschild solution
\eqref{C-Schwarzschild}, \eqref{Sch-drdu}.
This is realized in eq.\eqref{a=abar},
which relates $a_0(v)$ to $a(u)$ there.

\section{Relation between $a(u)$ and $\bar{a}(v)$}
\label{a-abar}
Here we derive the relation \eqref{a=abar}
between the Schwarzschild radii $a(u)$ and $\bar{a}(v)$
on the outer boundary of the near-horizon region.
Take the $v$-derivative of eq.\eqref{outer-boundary-condition},
which defines the location of the outer boundary of the near-horizon region,
we find
\begin{align}
\frac{\del r}{\del u} \frac{du_{out}(v)}{dv} + \frac{\del r}{\del v} - \frac{d\bar{a}}{dv}
= - \frac{N\ell_p^2}{\bar{a}^2(v)} \frac{d\bar{a}}{dv}.
\label{B2}
\end{align}
Use eqs.\eqref{C-Schwarzschild}, \eqref{Sch-drdu}, \eqref{outer-boundary-condition}
to estimate $\del r/\del u$ and $\del r/\del v$ as
\begin{align}
\frac{\del r}{\del u} \simeq -\frac{\del r}{\del v}
\simeq - \frac{1}{2}\left(1 - \frac{a}{r}\right) \simeq - \frac{N\ell_p^2}{2a^2}.
\end{align}
Then, together with eq.\eqref{dadu},
the equation above becomes
\begin{align}
\frac{N}{2}\left(1 - \frac{du_{out}(v)}{dv}\right) + \sigma \simeq \frac{N\sigma\ell_p^2}{a^2},
\end{align}
which implies that
\be
\frac{du_{out}(v)}{dv} \simeq 1 + \frac{2\sigma}{N},
\label{dudv-out}
\ee
assuming that $N \ll a^2/\ell_p^2$.

Next,
we take the $v$-derivative of $C(u_{out}(v), v)$ according to eq.\eqref{C-sol};
\begin{align}
\frac{d}{dv}C(u_{out}(v), v) &\simeq
\left[- \frac{1}{2a(u_{out}(v))}\frac{du_{out}(v)}{dv} + \frac{1}{2\bar{a}(v)}
- \frac{\del_v r(u_{out}(v), v)}{r(u_{out}(v), v)}\right]C(u_{out}(v), v),
\label{B5}
\end{align}
which should agree with the Schwarzschild approximation of the same quantity
\be
\frac{d}{dv}\left(1 - \frac{a(u_{out}(v))}{r(u_{out}(v), v)}\right)
\simeq \frac{d}{dv}\left(\frac{N\ell_p^2}{a^2(u_{out}(v))}\right)
\sim \mathcal{O}\left(\frac{\ell_p^4}{a^5}\right).
\ee
This agreement at the leading order of the $\ell_p^2/a^2$ expansion means
\be
\frac{a(u_{out}(v))}{\bar{a}(v)} \simeq
\frac{du_{out}(v)}{dv} \simeq 1 + \frac{2\sigma}{N}.
\ee
where we used eq.\eqref{B2} and
dropped the last term of eq.\eqref{B5} as a higher-order term.

\section{Order-of-magnitude of the first term in eq.\eqref{delur}}
\label{integral-evaluation}

Using eqs.\eqref{Tuu}, \eqref{C-sol}, and $r/a \sim \mathcal{O}(1)$,
the first term in eq.\eqref{delur} can be estimated as
\begin{align}
- \frac{\kappa}{2} C(u, v) \int_{u_{\ast}}^u du' \, \frac{r(u', v)}{C(u', v)}T_{uu}(u', v)
&\sim \mathcal{O}\left(\ell_p^2 C(u, v) \int_{u_{\ast}}^u du' \, C(u', v)\frac{1}{a^3}\right)
\nn \\
&\sim \mathcal{O}\left(\frac{\ell_p^2}{a^3} \, C(u, v) C(u_{\ast}, v) 
\int_{u_{\ast}}^u du' \, e^{-\int_{u_{\ast}}^{u'}\frac{du''}{2a(u'')}}\right),
\label{B-1}
\end{align}
where we assumed that the range $(u-u_{\ast}) \ll \mathcal{O}(a^3/\ell_p^2)$
so that the Schwarzschild radius $a$ remains the same order of magnitude. 
(This assumption is consistent with the range \eqref{uuvvrange}.) 
The integral above can then be estimated as
\begin{align}
\int_{u_{\ast}}^u du' \, e^{-\int_{u_{\ast}}^{u'}\frac{du''}{2a(u'')}}
\simeq \int_{u_{\ast}}^u du' \, e^{-\frac{u-u_{\ast}}{2a}}
\lesssim \mathcal{O}(a).
\end{align}

In the evaluation of eq.\eqref{delur},
we have taken $u_{\ast} = u_{out}(v)$
so that $(u_{\ast}, v)$ lies on the outer boundary of the near-horizon region.
Then we can use eqs.\eqref{C-Schwarzschild} and \eqref{outer-boundary-condition}
to evaluate $C(u_{out}(v), v) \simeq N\ell_p^2/a^2 \ll 1$.
Following eq.\eqref{B-1},
the first term in eq.\eqref{delur} is estimated as
\begin{align}
- \frac{\kappa}{2} C(u, v) \int_{u_{\ast}}^u du' \, \frac{r(u', v)}{C(u', v)}T_{uu}(u', v)
&\lesssim \mathcal{O}\left(\frac{\ell_p^2}{a^2} C(u, v) C(u_{\ast}, v)\right)
\nn \\
&\ll \mathcal{O}\left(\frac{\ell_p^2}{a^2} C(u, v)\right).
\end{align}
On the other hand,
the second term in eq.\eqref{delur} is of $\mathcal{O}(C)$.
Therefore,
the first term is negligible in comparison.

\section{Calculation of ${\cal M}_{\{1\cdots 1\}0}$}
\label{example-calc}

We evaluate M here by using the expression 
\eqref{Aml} for ${\cal A}_{\{1\cdots 1\}0}$:
\begin{align}
{\cal A}_{\{1\cdots 1\}0} =
\int_{\cal V} dU dV \,
r \left(\partial_V\frac{1}{r}\right)^n \, e^{i\om_U(U - U_h)}.
\label{UV-int-1}
\end{align}

\hide{
According to eqs.\eqref{U-u} and \eqref{V-v},
when the shell is deep inside the near-horizon region, 
$U \simeq U_h$ and $V = V_h + 2a$.
On the other hand,
eq.\eqref{r-sol-simpler} implies that
$R_s \simeq a$ deep inside the shell,
so we deduce from eq.\eqref{r-UV} that, 
when the shell is deep inside the near-horizon region,
$a \simeq R_s \simeq r_{in}(U_h, V_s) = \frac{(V_h + 2a) - U_h}{2} + \xi
= a + \frac{V_h - U_h}{2} + \xi$,
which implies eq.\eqref{xi}.
}
We consider the spacetime region ${\cal V}$ as shown in Fig.\ref{UV-1}.
Eq.\eqref{UV-int-1} includes all the contributions from the regions (i) -- (iv). 

As the areal radius $r$ has different functional forms inside and outside the shell,
the factor $\del_V(1/r)$ appearing in eq.\eqref{UV-int-1}
is of the following form
\begin{align}
\del_V\frac{1}{r} 
\simeq - \frac{\del_V r_{in}}{r^2} \; \Theta(V_s(U) - V)
- \frac{\del_V r_{out}}{r^2} \; \Theta(V - V_s(U)),
\label{r-step}
\end{align}
where $V_s(U)$ is the $V$-coordinate of the collapsing thin null shell
and $r_{in}$ ($r_{out}$) the areal radius inside (outside) the shell.
The step function $\Theta(V_s - V)$ selects the region inside the shell,
and $\Theta(V - V_s)$ that outside the shell.

In the flat space inside a collapsing shell, 
we have
\begin{align}
r = r_{in}(U, V) \equiv (V-U)/2 + \xi,
\label{r-UV}
\end{align}
where $\xi \equiv (U_h - V_h)/2$.
The value of $\xi$ is fixed by the continuity of $r$
across the thin shell when it is deep inside the near-horizon region,
using eqs.\eqref{U-u}, \eqref{V-v}, and \eqref{r-sol-simpler}.

According to eq.\eqref{r-UV}, 
$\del_V r_{in} = 1/2$.
We derive $\del_V r_{out}$ from
eqs.\eqref{a=abar}, \eqref{r-sol}, \eqref{dr0dv}, \eqref{U-u}, \eqref{V-v} and \eqref{C-UV} as
\begin{align}
\del_V r_{out} 
&\simeq
\left(\dot{\bar{a}}(v) + \frac{1}{2} C\right)\left(\frac{dV}{dv}\right)^{-1}
\simeq - \frac{\sigma\ell_p^2}{\bar{a}}
\left(\frac{2}{V - V_h}\right)
\left(1 - \frac{(U_h-U)(V - V_h)}{8\sigma\ell_p^2}\right)
\label{delVrout}
\end{align}
in the near-horizon region.
Using eq.\eqref{V-v},
we see that $V - V_h = 2a$ on the shell at $v = v_s$.
On the other hand,
$U_h - U$ becomes arbitrarily small deep inside the near-horizon region.

The step functions in eq.\eqref{r-step}
divide the integral \eqref{UV-int-1} into two parts:
\begin{align}
{\cal A}_{\{1\cdots 1\}0} = 
{\cal A}^{(inside)}_{\{1\cdots 1\}0} + {\cal A}^{(outside)}_{\{1\cdots 1\}0}.
\label{AAA}
\end{align}
${\cal A}^{(inside)}_{\{1\cdots 1\}0}$ is the contribution from the region (iv),
and ${\cal A}^{(outside)}_{\{1\cdots 1\}0}$ is that from the regions (i) and (ii).
Note that there is no contribution from (iii) due to the absence of $\delta (V-V_s)$ in eq.\eqref{r-step}.  

Before evaluating the contributions inside and outside the collapsing shell
to the transition amplitude,
we note that the spacetime is divided into two parts here as
${\cal A}^{(inside)}_{\{1\cdots 1\}0}$ and ${\cal A}^{(outside)}_{\{1\cdots 1\}0}$
by a physical object --- the null shell.
This is in contrast with the calculation of matrix elements in which
the spacetime is divided into two parts by the event horizon.
Since the event horizon has no local physical meaning,
it was found in Ref.\cite{Giddings:2006sj} that 
the contributions of the two parts of the spacetime cancel to a large extent 
in the calculation of certain matrix elements.

On the other hand,
in the near-horizon region,
it is unlikely to have generic cancellation between
${\cal A}^{(inside)}_{\{1\cdots 1\}0}$ and ${\cal A}^{(outside)}_{\{1\cdots 1\}0}$
because only the region outside the shell depends on the mass.
As we will see below,
the large difference between $\del_V r_{in}$ and $\del_V r_{out}$
across the null shell in the near-horizon region
leads to a significant contribution to the amplitude ${\cal A}_{\{1\cdots1\}0}$.


To define a complete basis of functions in this region,
we impose the periodic boundary conditions in $T$ for convenience.
($T$ is defined in eq.\eqref{T-def}.)
The frequency $\om_U$ is thus quantized as
\begin{align}
\om_U \in \frac{2\pi \mathbb{Z}}{T_1 - T_0}.
\label{quant-omU}
\end{align}
The integral over ${\cal V}$ can be easily carried out
using the following formula
\begin{align}
\int_{x_0}^{x_1} dx \, f(x) e^{i\om_U x} \simeq
\frac{f(x_1) \, e^{i\om_U x_1} - f(x_0) \, e^{i\om_U x_0}}{i\om_U},
\label{int-id}
\end{align}
where we assumed that
\begin{align}
\left|\frac{f''(x)}{f'(x)}\right|, \left|\frac{f'''(x)}{f'(x)}\right|^{1/2},
\cdots
\ll \om_U.
\end{align}
This will be a good approximation because
the integral over $\om_U$ will be dominated by 
a trans-Planckian value $\sim \om (dU/du)^{-1}$
with $\om \sim 1/a$ \eqref{om-others}.
We will apply this formula \eqref{int-id} to integrals over 
the variables $V$ and $T$ below.

The shell is collapsing at the speed of light at $V = V_s$,
with the areal radius
\be
r = R_s(T) \equiv V_s - T + \xi,
\label{r=V-T}
\ee
assuming that
$R_s(T) \in (r_0, r_1)$ for $T \in (T_0, T_1)$.
Now, we evaluate ${\cal A}^{inside}$ using
eqs.\eqref{UV-int-1}, \eqref{r-step}, \eqref{r-UV}, \eqref{int-id},
we find
\begin{align}
{\cal A}^{(inside)}_{\{1\cdots 1\}0}
&= 2\int_{T_0}^{T_1} dT \int_{T + r_0-\xi}^{V_s} dV \,
r \left(\frac{-1}{2r^2}\right)^n \, e^{i\om_U(2T - V - U_h)}
\nn \\
&\simeq 2 \left(\frac{-1}{2}\right)^n 
\int_{T_0}^{T_1} dT
\frac{1}{-i\om_U}\left[
\frac{1}{R_s^{2n-1}(T)} \, e^{i\om_U(2T - V_s - U_h)}
- \frac{1}{r_0^{2n-1}} \, e^{i\om_U(T - r_0 - U_h + \xi)}
\right].
\nn
\end{align}
Note that the 2nd term in the integral on the right-hand side
has no contribution due to the condition \eqref{quant-omU}.
Hence, using eqs.\eqref{quant-omU} and \eqref{int-id} again, we obtain
\begin{align}
{\cal A}^{(inside)}_{\{1\cdots 1\}0}
&\simeq 2 \left(\frac{-1}{2}\right)^n 
\frac{1}{2\om_U^2} 
\left[
\frac{1}{R_s^{2n-1}(T_1)}
- \frac{1}{R_s^{2n-1}(T_0)}
\right] \, e^{i\om_U(2T_1 - V_s - U_h)}
\nn \\
&\simeq \left(\frac{-1}{2}\right)^n 
\frac{1}{\om_U^2} 
\frac{(2n-1)}{R_s^{2n}(T_0)} \, (T_1 - T_0)
\, e^{i\om_U(2T_1 - V_s - U_h)}
\nn \\
&\sim \left(\frac{-1}{2}\right)^n 
\frac{(2n-1)}{a^{2n}} \frac{(T_1 - T_0)}{\om_U^2}
\, e^{i\om_U(2T_1 - V_s - U_h)},
\end{align}
where in the 2nd last line we used 
\be
\frac{R_s^{-(2n-1)}(T_1) - R_s^{-(2n-1)}(T_0)}{T_1-T_0}
\approx \left.\frac{dR_s^{-2n+1}}{dT}\right|_{T_0}
= - \frac{(2n-1)}{R_s^{2n}(T_0)}
\label{RdR}
\ee
for $T_1-T_0 \ll a$,
and in the last line we have used $R_s(T_0) ~ a$
as the typical order of magnitude on the shell.

Similarly,
letting $V_1(T)$ denote the upper bound of the $V$-integration
corresponding to $r = r_1 \gg a$,
we have, for $n > 2$,
\begin{align}
{\cal A}^{(outside)}_{\{1\cdots 1\}0}
&\simeq
2\int_{T_0}^{T_1} dT \int_{V_s}^{V_1(T)} dV \,
r \left(- \frac{\del_V r}{r^2}\right)^n
\, e^{i\om_U(2T - V - U_h)}
\nn \\
&\sim 2\int_{T_0}^{T_1} dT \,
\frac{\bar{a}}{i\om_U} \left(- \frac{\sigma\ell_p^2}{\bar{a}^3}\right)^n
\left(\frac{2}{V_s - V_h}\right)^n
\left(1 - \frac{n (U_h-2T+V_s)(V_s-V_h)}{8\sigma\ell_p^2}\right)
\, e^{i\om_U(2T - V_s - U_h)},
\nn
\end{align}
where $R_s$ is replaced by $\bar{a}$ as an order-of-magnitude estimate,
and we have used eq.\eqref{delVrout} 
(and the Taylor expansion of its $n$-th power)
as well as eq.\eqref{int-id}.
Here, the spacetime at $V=V_1(T)$ is far away the near-horizon region, 
and the contribution is negligible due to $r=r_1 \gg a$ compared to that from $V=V_s$. 
The spacetime at $V=V_s$ is inside the near-horizon region, 
and eq.\eqref{delVrout} has been used. 
Using eqs.\eqref{quant-omU} and \eqref{int-id} again for the integration over $T$, 
we find
\begin{align}
{\cal A}^{(outside)}_{\{1\cdots 1\}0}
&\sim -
\frac{\bar{a}}{\om_U^2}
\left(- \frac{\sigma\ell_p^2}{\bar{a}^3}\right)^n
\left(\frac{2}{V_s - V_h}\right)^n 
\Big[
\left(1 - \frac{n (U_h-2T+V_s)(V_s-V_h)}{8\sigma\ell_p^2}\right)
\, e^{i\om_U(2T - V_s - U_h)}
\Big]_{T_0}^{T_1}
\nn \\
&\sim 
(-1)^{n-1} \frac{n \sigma^{n-1}\ell_p^{2(n-1)}}{2 \bar{a}^{4n-2}}
\frac{(T_1-T_0)}{\om_U^2} \, e^{i\om_U(2T_1 - V_s - U_h)},
\end{align}
where we used $V_s - V_h = 2a$ according to eq.\eqref{V-v}.
\footnote{
We can use eq.\eqref{Sch-drdu} to derive
$
\del_V r_{out} \simeq \left(\frac{dV}{dv}\right)^{-1} \frac{\del r}{\del v}
\simeq \frac{1}{2} \left(1 - \frac{a}{r}\right),
$
where eq.\eqref{V-v} is used to deduce $dV/dv = 1$ at $v = v_s$.
}

Thus,
${\cal A}^{(outside)}_{\{1\cdots 1\}0}$ is negligible
in comparison with ${\cal A}^{(inside)}_{\{1\cdots 1\}0}$.
The origin of this hierarchy is the large difference in $\del_V r$
inside and outside the shell mentioned above.
If the shell is not in the near-horizon region, 
but far away from the horizon $(r \gg a)$,
${\cal A}^{(outside)}_{\{1\cdots 1\}0}$ and ${\cal A}^{(inside)}_{\{1\cdots 1\}0}$
would be of the same order of magnitude
and have the possibility of a large cancellation between them.

Plugging ${\cal A}_{\{1\cdots 1\}0}$ back into eq.\eqref{UV-integral},
the integral $\int d\om_U$ should be replaced by the sum
over $\om_U = 2\pi m/(T_1 - T_0)$ with $m \in \mathbb{Z}_+$ as
\begin{align}
\int_0^{\infty} d\om_U \, \om_U^{n - 3 - i2a\om} e^{-i\om_U(U_h - U_s)} 
&\rightarrow
\sum_{m = 1}^{\infty} \frac{2\pi}{(T_1 - T_0)} \left(\frac{2\pi m}{T_1 - T_0}\right)^{n-3-i2a\om} \;
e^{-i\frac{2\pi m}{T_1 - T_0}(U_h - U_s)}
\nn \\
&= \left(\frac{2\pi}{T_1 - T_0}\right)^{n-2-i2a\om}
\mbox{PolyLog}(-(n-3-i2a\om), e^{-i\frac{2\pi}{T_1 - T_0}(U_h - U_s)})
\nn \\
&\sim
\zeta(-(n-3-i2a\om))\Gamma(n-2-i2a\om) (U_h - U_s)^{-(n-2-i2a\om)},
\end{align}
where $U_s \equiv 2T_1 - V$ is the $U$-coordinate of the collapsing shell at $T_1$.
In the expression above,
we have assumed that $(T_1 - T_0) \gg (U_h - U_s) = 2aC(u_s, v_s)$.
This is consistent with the consideration of a scrambling time for a distant observer.

Using the identity
\begin{align}
\Gamma(ib)\Gamma(-ib) = \frac{\pi}{b\sinh(\pi b)}
\end{align}
and
\begin{align}
C(u_s, v_s) = \frac{dU}{du}(u_s) \simeq \frac{U_h - U_s}{2a},
\end{align}
where $u_s$ is the $u$-coordinate of the point $(T = T_1, r = R_s(T_1))$,
the transition amplitude \eqref{UV-integral} is found to be
\begin{align}
{\cal M}_{\{1\cdots 1\}0}
&\sim
\frac{4 (2n-1) \lam_{\{1\cdots 1\}0} a\om
\Gamma(i2a\om)\Gamma(n-2-i2a\om) e^{-\pi a\om} \zeta(-(n-3-i2a\om))
\ell_p^{3n-3}}{a^{2n}} \,
\frac{(T_1 - T_0)}{(U_h - U_s)^{n-2}}
\nn \\
&\sim
\frac{2\pi (2n-1) (n-3)! \lam_{\{1\cdots 1\}0} e^{-\pi a\om} \zeta(-(n-3-i2a\om))
\ell_p^{3n-3}}{2^{n-2} \sinh(2\pi a\om) a^{3n-3}} \,
\frac{(T_1 - T_0)}{a} \;
C^{-(n-2)}(u_s, v_s)
\label{M1110}
\end{align}
up to a factor of $\mathcal{O}(1)$.

One might suspect that the origin of the large amplitude
is the $\delta$-function energy density of the thin shell.
A shell with a smooth energy density
could in principle lead to a smaller ${\cal A}^{(inside)}_{\{1\cdots 1\}0}$,
but, as mentioned above,
even the contribution of the vacuum energy
is sufficient to induce a large amplitude within the scrambling time.
The conclusion is robust because of the exponential behavior of $C(u, v)$.

\end{document}